\documentclass[11pt]{article}
\usepackage[left=1in,top=2cm,right=1in,nohead]{geometry}
\usepackage[dvipdf]{graphicx}
\usepackage{amssymb,amsfonts,amsmath}

\usepackage{color}
\usepackage{multirow}
\usepackage[colorlinks,citecolor={black},urlcolor={black},linkcolor={black}]{hyperref}

\newcommand{\stb}[1]{{{\hbox{\sc{st}}(#1)}}}
\newcommand{\ctb}[1]{{{\hbox{\sc{c}}(#1)}}}
\newtheorem{theorem}{Theorem}[section]
\newtheorem{lemma}[theorem]{Lemma}

\newcommand{\weburl}[1]{{\url{#1}}}

\title{A single-photon sampling architecture for \\ solid-state imaging
  sensors}

\author{Ewout van den Berg\footnote{Department of Mathematics,
    Stanford University, Stanford CA 94305}, Emmanuel
  Cand\`{e}s\footnote{Departments of Mathematics and Statistics,
    Stanford University, Stanford CA 94305}, Garry
  Chinn,\!\!\footnote{Department of Radiology, Stanford University,
    Stanford CA 94305} \\Craig Levin\footnote{Departments of
    Radiology, Physics, Electrical Engineering, and Bioengineering,
    Stanford University, Stanford \ \hspace*{17pt}CA 94305}, Peter
  Olcott\footnote{Departments of Radiology and Bioengineering,
    Stanford University, Stanford CA 94305}, and Carlos
  Sing-Long\footnote{Institute for Computational and Mathematical
    Engineering, Stanford University, Stanford CA 94305}}

\begin{document}

\maketitle

\begin{abstract}
Advances in solid-state technology have enabled the development of
silicon photomultiplier sensor arrays capable of sensing individual
photons. Combined with high-frequency time-to-digital converters
(TDCs), this technology opens up the prospect of sensors capable of
  recording with high accuracy both the time and location of each
  detected photon. Such a capability could lead to significant
  improvements in imaging accuracy, especially for applications
  operating with low photon fluxes such as LiDAR and positron emission
  tomography.

  The demands placed on on-chip readout circuitry imposes stringent
  trade-offs between fill factor and spatio-temporal resolution,
  causing many contemporary designs to severely underutilize the
  technology's full potential. Concentrating on the low photon flux
  setting, this paper leverages results from group testing and
  proposes an architecture for a highly efficient readout of pixels
  using only a small number of TDCs, thereby also reducing both cost and
  power consumption. The design relies on a multiplexing technique
  based on binary interconnection matrices. We provide optimized
  instances of these matrices for various sensor parameters and give
  explicit upper and lower bounds on the number of TDCs required to
  uniquely decode a given maximum number of simultaneous photon
  arrivals.

  To illustrate the strength of the proposed architecture, we note a
  typical digitization result of a 120 $\times$ 120 photodiode sensor
  on a 30 $\mu$m $\times$ 30 $\mu$m pitch with a 40 ps time resolution
  and an estimated fill factor of approximately 70\%, using only 161
  TDCs. The design guarantees registration and recovery of up to $s=4$
  simultaneous photon arrivals. A fast decoding algorithm is
  available, which decodes successfully with probability $1-\alpha_s$
  whenever $s> 4$. By contrast, a cross-strip design requires 240 TDCs
  and cannot uniquely decode any simultaneous photon arrivals.  In a
  series of realistic simulations of scintillation events in clinical
  positron emission tomography the design was able to recover the
  spatio-temporal location of 98.6\% of all photons that caused pixel
  firings.
\end{abstract}



\section{Introduction}

Photon detection has become an essential measurement technique in
science and engineering. Indeed, applications such as positron
emission tomography (PET), single-photon emission computed tomography,
flow cytometry, LiDAR, fluorescence detection, confocal microscopy,
and radiation detection all rely on accurate measurement of photon
fluxes. Traditionally, the preferred measurement device has been the
photomultiplier tube (PMT), which is a high-gain, low-noise photon
detector with a high-frequency response and a large area of
collection.  In particular, it behaves as an ideal current generator
for steady photon fluxes, making it suitable for use in applications
in astronomy and medical imaging, among others. However, they are
bulky, require manual assembly steps, and have limited spatial
resolution for PET. For these reasons, extensive research has focused
on finding feasible solid-state alternatives that can operate at much
lower voltages, are immune to magnetic effects, have increased
efficiency, and are smaller in physical size~\cite{REN2006a}. Recent
designs have raised significant interest in the community, and their
use as a replacement for PMTs in applications such as PET
imaging~\cite{HEN2009GZSa}, high-energy physics~\cite{SEF2007a},
astrophysics~\cite{TES2007DMNa}, LiDAR~\cite{AUL2002LYHa}, and flow
cytometry has been recently explored.


Silicon photomultiplier (SiPM) devices consist of two-dimensional
arrays of Geiger avalanche photodiodes (APD) that are run above their
breakdown voltage and are integrated with either an active or passive
quenching circuit. Devices built up from these Geiger APD microcells
are characterized by the fraction of sensing area on the device (the
fill factor), and the fraction of incident photons that cause charged
APDs to fire (the detection quantum efficiency). The product of these
two quantities gives the combined photon detection efficiency
(PDE).

The compatibility of Geiger APDs with standard CMOS technology has
enabled a number of different designs. For example, the digital
silicon photomultiplier (DSiPM) \cite{FRA2009PDGa} adds processing
circuitry to count the total number of photons hitting the sensor and
records the time stamp of the first group of photons that crosses a
predetermined photon intensity threshold. The resolution of the
resulting sensor coincides with the size of the entire sensor (usually
3 mm $\times$ 3 mm in area) whereas the temporal sampling is limited
to a single time stamp per pulse.  Nevertheless, the sensor does
achieve a very high fill factor (80\%) and is, therefore, very
sensitive. Note that in SiPM terminology, the above tiling of
microcells into larger atomic units is often referred to as a
pixel. Throughout this paper, however, we use the term pixel to
refer to individual Geiger APD microcells, since they represent the
smallest resolvable element in our proposed design.

An alternative sensor design, called SPAD, aims at a high temporal
resolution and registers the time of each pixel (APD) firing
\cite{ISA2010PBHa,MAR2010TZTa}. This is achieved by connecting each
pixel to a high-performance time-to-digital converter (TDC), which
records a time stamp in a memory buffer whenever a signal is detected
on its input. Because of their relatively low complexity, especially
when compared to analog-to-digital converters, TDCs allow the sensor
to achieve an extremely high temporal resolution. However, the spatial
resolution of the sensor is severely compromised by the large amount
of support circuitry between neighboring pixels, resulting in an
extremely low fill factor of approximately 1--5\%.

\begin{figure}[t*]
\centering
\begin{tabular}{cccc}
\includegraphics[height=3.95cm]{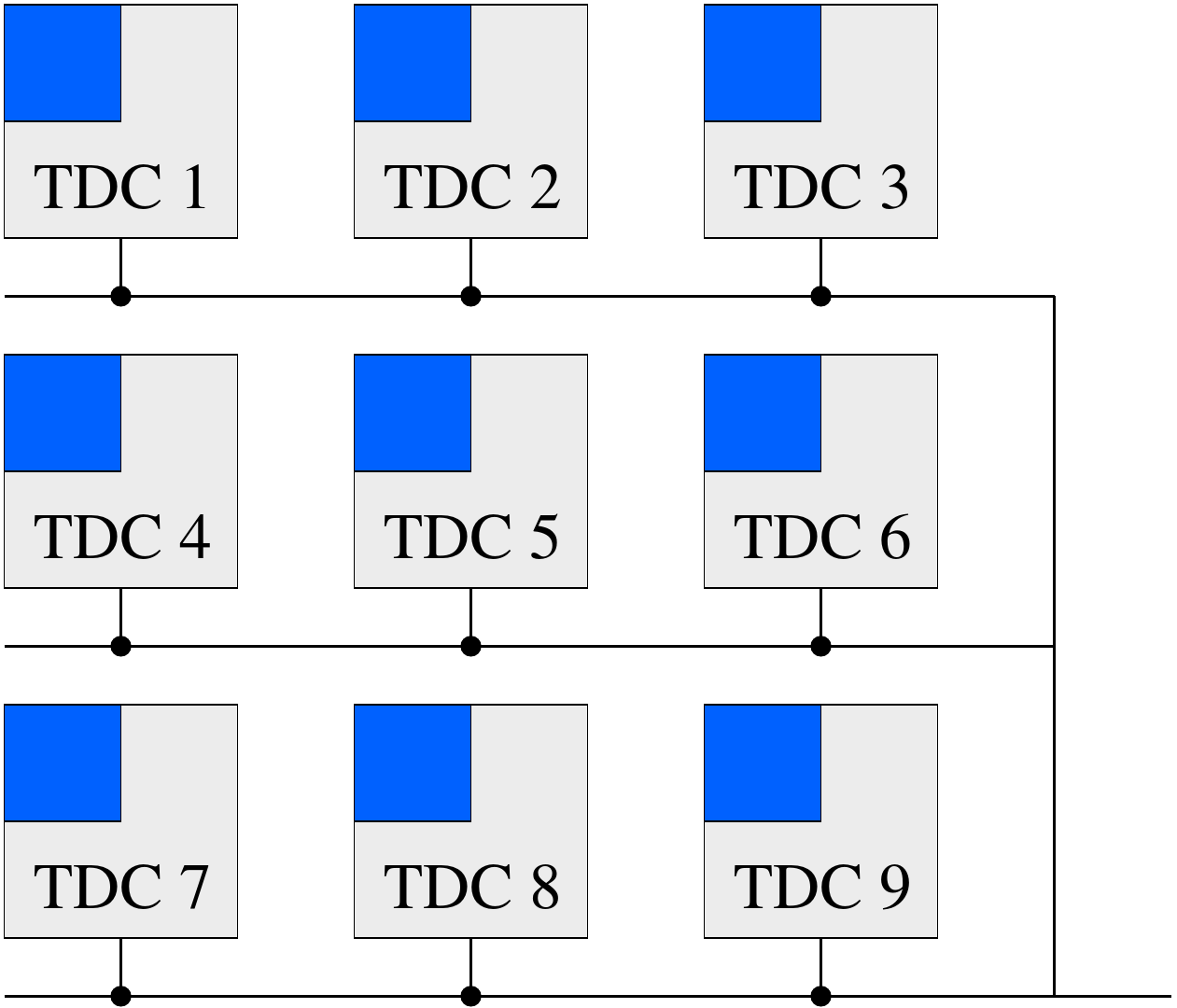}&
\includegraphics[height=3.95cm]{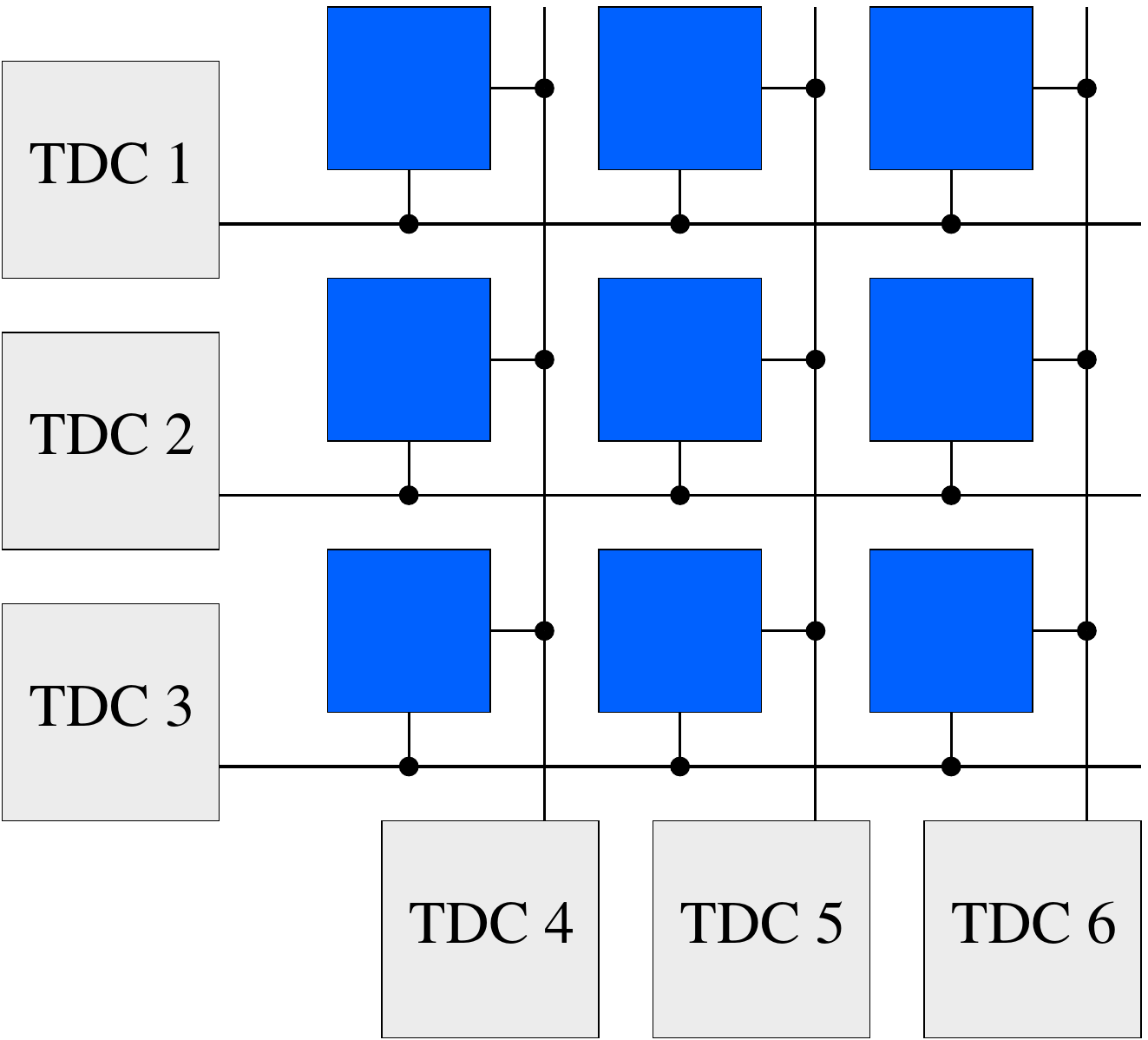}&&
\includegraphics[height=3.95cm]{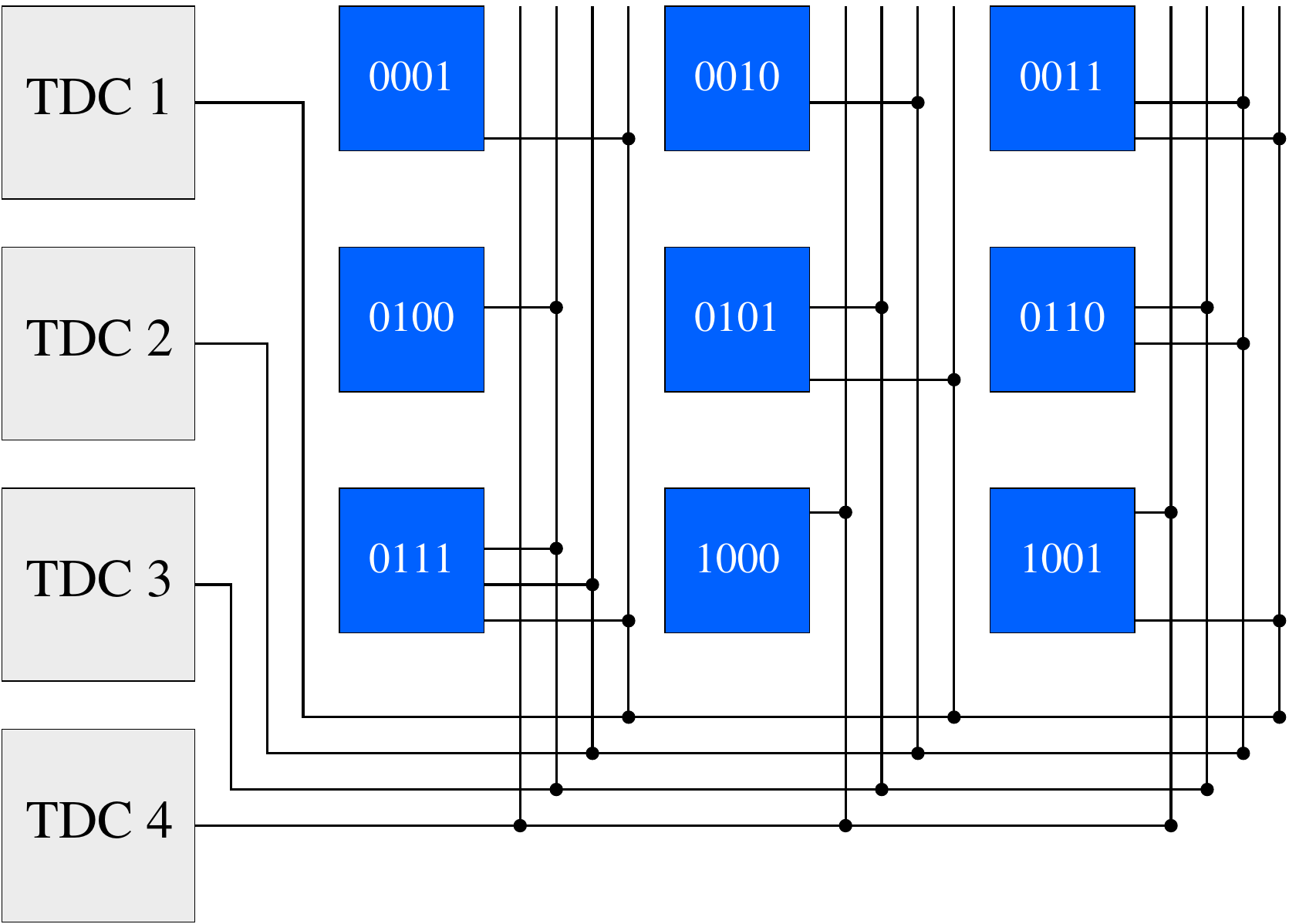}\\
(a) & (b) && (c)
\end{tabular}
\caption{Three SiPM designs with (a) one TDC per pixel, (b) one TDC per
  row and column of pixels, and (c) one TDC per bit of the binary
  representation of each pixel number. The first design is capable of
  detecting any number of simultaneous hits, at the cost of a large
  number of TDCs. The other two can only uniquely decode up to a
  single pixel firing, but require substantially fewer
  TDCs. }\label{Fig:Design}
\end{figure}
 
Although promising, both designs show that current implementations
highly underutilize the full potential of these silicon devices: due
to the restricted chip area a trade-off must be made between the
spatio-temporal resolution and the fill factor. In order to improve
this trade-off, we take advantage of the special properties present in
settings with a low photon flux. In particular, we claim that, by
taking advantage of the temporal sparsity of the photon arrivals, we
can increase the spatio-temporal resolution of the SiPM to within a
fraction of the theoretical maximum while maintaining simple circuitry
and a high fill factor.

This claim is made possible by the combination of two ideas: (1) to
use TDCs as the main readout devices, and (2) to exploit ideas from
\emph{group testing} to reduce the number of these devices. As a
motivational example, consider three possible designs for an $m\times
m$ pixel grid. The first design, illustrated in
Figure~\ref{Fig:Design}(a) corresponds to a design with a single TDC
per pixel.  This design can be seen as a trivial group test capable of
detecting an arbitrary number of simultaneous firings, but at the cost
of $n^2$ TDCs. When the photon flux is low, only a small number of
pixels will typically fire at the same time\footnote{Throughout the
  paper, firing at the same time is to be understood as within the
  same TDC sampling interval.}, thus causing most TDCs to be idle most
of the time, and resulting in a very poor usage of resources. The
second design, shown in Figure~\ref{Fig:Design}(b), corresponds to the
widely-used cross-strip architecture in which rows and columns of
pixels are each aggregated into a single signal. This design requires
only $2n$ TDCs, but information from the TDCs can only be uniquely
decoded if no more than one pixel fired during the same time
interval. Hinting at the power of more efficient group-testing
designs, Figure~\ref{Fig:Design}(c) shows a design in which pixels are
numbered from 1 to $n^2$ and in which a pixel $i$ is connected to TDC
$j$ only if the $j$-th bit in the binary representation number $i$ is
one. This design is also capable of decoding up to a single pixel
firing, but requires only $\lfloor \log_2 2n\rfloor$ TDCs. The
objective of this paper is to show that whenever the number of
simultaneous firings is small, a significant reduction in the number
of TDCs (and accompanying memory buffers) can be attained by carefully
selected designs, which reduce both the amount of overhead circuitry
and the generated volume of data.

\subsection{Contributions and paper organization}

The main contribution of this paper are: (1) the introduction of group
testing to the design of interconnection networks in imaging sensors,
combined with TDC readout, (2) an architecture to time and locate
photon arrivals, (3) the construction of highly optimized group testing
matrices for a variety of conventional $m\times m$ grid sizes, (4) a
comparison with alternative constructions, and (5) an extensive
comparison with explicitly evaluated and optimized theoretical upper
and lower bounds on the minimum number of TDCs required to decode an
imaging sensor array.

The paper is organized as follows. We start by reviewing the concept
of group testing and derive the currently best-known group testing
designs for a range of grid sizes. We then provide an extensive survey
of constructions used to obtain upper and lower bound on the minimum
number of TDCs needed to guarantee recovery for $d=2,\ldots,6$
simultaneous hits. There, the results show that the designs generated
earlier are close to the theoretical lower bounds. We then test the
performance of the design on simulated data generated using a
realistic model of scintillation events arising in PET scanners. We
conclude with a discussion on practical considerations in the
implementation of the proposed design.
  
Throughout the paper we use the following notation: $\log$ is base $e$
unless otherwise indicated; $[n]$ denotes the set $\{1,\ldots,n\}$;
and ${[n] \choose k}$ denotes the family of all subsets of $[n]$ with
$k$ elements.

\section{Group testing}

Group testing was proposed by Dorfman \cite{DOR1943a} to effectively
screen large numbers of blood samples for rare diseases (see
\cite{DU2000Ha} for more historical background). Instead of testing
each sample individually, carefully chosen subsets of samples are
pooled together and tested as a whole. If the test is negative, we
immediately know that none of the samples in the pool is positive,
saving a potentially large number of tests. Otherwise, one or more
samples is positive and additional tests are needed to determine
exactly which ones.

Since its introduction, group testing has been used in a variety of
different applications including quality control and product testing
\cite{SOB1959Ga}, pattern matching \cite{CLI2007EPRa,IND1997a}, DNA
library screening \cite{BAL1996BKTa,DU2006Ha,DYA2000MRb,NGO2000Da},
multi-access communication \cite{WOL1985a}, tracking of hot items in
databases \cite{COR2006Ma}, and many others \cite{DU2000Ha,POR2008Ra}.
Depending on the application, group testing can be applied in an
adaptive fashion in which tests are designed based on the outcome of
previous tests, and in a nonadaptive fashion in which the tests are
fixed {\it a priori}. Additional variations include schemes that
provide robustness against test errors \cite{KNI1998BTa,MAC1997a}, or
the presence of inhibitors \cite{DEB2005GVa,FAR1997KKMa}, which cause
false positives and negatives, respectively.

In our SiPM application, each pixel fires independently\footnote{This
  assumes that pixels have been shielded to avoid cross-talk.}
according to some Poisson process, and exactly fits in the
probabilistic group-testing model \cite{BAL1996BKTa}. Nevertheless,
since we are interested in guaranteeing performance up to a certain
level of simultaneous firings, we will study the application from a
combinatorial group-testing perspective. Furthermore, measurements are
necessarily nonadaptive as they are hardwired in the actual
implementation. No error correction is needed; the only errors we
can expect are spurious pixel firings (dark counts) or afterpulses,
but these appear indistinguishable from real firings, and cannot be
corrected for.  The rate of these spurious pixel firings is usually
much less than the signal rate, especially in our scintillation
example \cite{CAS2009SVa}.

\subsection{Matrix representation and guarantees}

A group test can be represented as a binary $t\times n$ incidence
matrix or code $A$ with $t$ the length or number of tests and $n$ the
size or number of items, or pixels in our case. An entry $A_{i,j}$ is
set to one if item $j$ is pooled in test $i$, that is, if pixel $j$ is
connected to TDC $i$, and zero otherwise. Given a vector $x$ of length
$n$ with $x_j = 1$ if item $j$ is positive and zero otherwise, we can
define the test vector $y$ as $y = Ax$, where multiplication and
addition are defined as logic {\sc and} and {\sc or},
respectively\footnote{Addition using the logic {\sc{or}} gives $0+0=0$
  and $0+1 = 1+1 = 1$, while $0\cdot 0 = 0 \cdot  1 = 0$ and $1 \cdot 1=1$, as
  usual.}. The columns $a_j$ in $A$ are called codewords, and for sake
of convenience we allow set operations to be applied to these binary
codewords, acting on or defining their support. As an illustration of
the above, consider the following example:
\begin{center}
\begin{tabular}{cccc}
& & \scriptsize$\begin{array}{ccccc}\mathit{1}&\mathit{2}&\mathit{3}&\mathit{4}&\mathit{5}\end{array}$ & \multirow{2}{*}{$\left(\begin{array}{c}
0 \\ 1 \\ 0 \\ 1 \\ 0
\end{array}\right)$} \\
$\left(\begin{array}{c}1 \\ 1 \\ 0 \end{array}\right)$ &
= &
$\left(\begin{array}{ccccc}
1 & 0 & 0 & 1 & 1\\
0 & 1 & 0 & 1 & 0 \\
1 & 0 & 1 & 0 & 0
\end{array}\right)$ &
\\
\\
$y$ & = & $A$ & $x$
\end{tabular}
\end{center}
In this example pixels 1, 4, and 5 are connected to the first TDC, as
represented by the first row of $A$. When pixels 2 and 4 fire, we sum
the corresponding columns and find that the first and second TDCs are
activated, as indicated by $y$.

For group testing to be effective we want to have far fewer tests than
items ($t\ll n$). This inherently means that not all vectors $x$ can
be reconstructed from $y$, so we will be interested in conditions that
guarantee recovery when $x$ has up to $d$ positive entries.

The weakest notion for recovery is $d$-separability, which means that
no combination of exactly $d$ codewords can be expressed using any
other combination of $d$ codewords. A matrix is said to be
$\bar{d}$-separable if the combination of any up to $d$ codewords is
unique. This immediately gives an information-theoretic lower bound on
the length $t$:
\begin{equation}\label{Eq:InfThLowerBound}
t \geq \log_2\left( \sum_{i=0}^d {n \choose i}\right).
\end{equation}
When $d$ is small compared to $n$, this gives $t\approx d
\log_2(n/d)$.  A stronger criteria is given by $d$-disjunctness. Given
any two codewords $u$ and $v$, we say that $u$ covers $v$ if $u \cup v
= u$. Based on this, define $A$ to be $d$-disjunct if no codeword in
$A$ is covered by the union of any $d$ others. The concept of
disjunctness of sets and codes has been proposed in a number of
settings and such codes are also known as superimposed codes
\cite{KAU1964Sa}, or $d$-cover free families \cite{ERD1985FFa}.
The advantage of $d$-disjunct codes is that the positive entries in
any $d$-sparse vector $x$ correspond exactly to those codewords in $A$
that remain after discarding all codewords that have a one in a
position where $y$ is zero. This is unlike general separable matrices,
where a combinatorial search may be needed for the decoding.

The central goal in group testing is to determine the smallest length
$t$ for which a matrix of size $n$ can satisfy certain separability or
disjunctness properties. Denote by $S(n,d)$ and $D(n,d$) the set of
all $t\times n$ matrices $A$ that are $d$-separable, respectively
$d$-disjunct. Then we can define $T_{S}(n,d)$ the smallest length of
any $A \in S(n,d)$, and likewise for $T_{D}(n,d)$. As a further
refinement in classification we define $D_w(n,d)$ as all
constant-weight $d$-disjunct matrices of size $n$, i.e., those
matrices whose columns all have the same weight or number of nonzero
entries (note that the weights of any two matrices within this
class can differ). The overlap between any two columns $a_i$ and $a_j$
of a code is defined as $\vert a_i \cap a_j\vert$. Consider a
constant-weight $w$ code $A$ with maximum pairwise overlap
\[
\mu = \mu(A) := \max_{i\neq j} \vert a_i \cap a_j\vert.
\]
It is easily seen that it takes $\lceil w / \mu\rceil$ columns to cover
another one, and therefore that
\begin{equation}\label{Eq:LowerBoundDisjunctness}
d \geq \left\lfloor \frac{w-1}{\mu}\right\rfloor.
\end{equation}
We can then define $D_{w,\mu}(n,d)$ as the class of constant-weight
codes of size $n$ where the right-hand side of
\eqref{Eq:LowerBoundDisjunctness} equals $d$. As for all other
classes, the actual disjunctness of the codes in $D_{w,\mu}$ may still
be higher, however, there is a subtle distinction in parameterization
here in that, by definition, we have $D_{w,\mu}(n,d+1) \cap
D_{w,\mu}(n,d) = \emptyset$, whereas $D_w(n,d+1) \subset D_w(n,d)$,
and likewise for $D$ and $S$. Summarizing, we have
\[
D_{w,\mu}(n,d) \subseteq D_w(n,d) \subseteq D(n,d) \subseteq S(n,d),
\]
along with the associated minimum lengths $T(n,d)$.

\subsection{Matrix construction}

In this section we discuss three methods for creating $d$-disjunct
binary matrices. In addition, we construct matrices for the particular
case where $n=3,\!600$, corresponding to a $60\times 60$ pixel array.

\subsubsection{Greedy approach}

The greedy approach generates $d$-disjunct matrices one column at a
time. For the construction of a constant-weight $w$ matrix of length
$t$ the algorithms starts with $\mathcal{F} = \emptyset$, and proceeds
by repeatedly drawing, uniformly at random, an element $F$ from ${[t]
  \choose w}$ as a candidate column. Whenever the distance to all sets
already in $\mathcal{F}$ exceeds $w/d$ we add $F$ to $\mathcal{F}$ and
continue the algorithm with the updated family. This process continues
until either $\vert \mathcal{F}\vert = n$ or a given amount of time
has passed.

We applied the greedy algorithm to construct $d$-disjunct matrices of
size at least $3,\!600$. For each value of $d$ and length $t$, the
algorithm was allowed to run for 12 hours. Row (l) of
Table~\ref{Table:BoundsOnTnd} gives the minimum $t$ for which the
algorithm found a solution. Instances with fewer rows are very well
possible; this strongly depend both on the choice of the initial few
columns and the amount of time available.

A further reduction in the code length could be achieved by checking
disjunctness in a groupwise, rather than a pairwise setting. However,
the number of possible $d$-subsets of columns to consider at each
iteration grows exponentially in $d$, thereby rendering this approach
intractable for even moderate values of $d$.

\subsubsection{Chinese remainder sieve}\label{Sec:EPP2007GHa}

Eppstein et al.~\cite{EPP2007GHa} recently proposed the `Chinese
remainder sieve', a number-theoretic method for the deterministic
construction of $d$-disjunct matrices. In order to generate a
$d$-disjunct matrix with at least $n$ columns, first choose a sequence
$\{p_1^{e_1}, p_2^{e_2},\ldots, p_k^{e_k}\}$ of powers of distinct
primes such that $\prod_{i=1}^k p_i^{e_i} \geq n^d$. A $t\times n$
matrix $A$ with $t = \sum_{i=1}^k p_i^{e_i}$ is then created by
vertical concatenation of $p_{\ell}^{e_{\ell}}\times n$ matrices
$A_{\ell}$ as follows:
\[
A = \left[\begin{array}{c}A_1\\ A_2 \\ \vdots \\
    A_k\end{array}\right],
\qquad
A_{\ell}(i,j) = \begin{cases} 1 & i \equiv j \mod p_{\ell}^{e_{\ell}} \\
0 & \mathrm{otherwise}. \end{cases}
\]
As an example, consider the construction of a $1$-disjunct $5\times 6$
matrix generated based on $p_1 = 2$ and $p_2 = 3$. Applying the above
definition we obtain:
\[
A = \left[\begin{array}{c} \multirow{2}{*}{$A_1$} \\ \\ \hline \\
    A_2 \\ \\ \end{array}\right] =
\left[\begin{array}{cccccc}
1 & 0 & 1 & 0 & 1 & 0 \\
0 & 1 & 0 & 1 & 0 & 1\\
\hline
1 & 0 & 0 & 1 & 0 & 0 \\
0 & 1 & 0 & 0 & 1 & 0 \\
0 & 0 & 1 & 0 & 0 & 1 
\end{array}\right].
\]
The above construction requires $t = \mathcal{O}(d^2 \log^2 n / (\log
d + \log\log n))$ rows for a $d$-disjunct matrix of size $n$
\cite{EPP2007GHa}. This is only slightly worse than the best-known
$\mathcal{O}(d^2\log n)$ construction, which we discuss in more detail
in Section~\ref{Sec:AsymptoticBounds}. Using a near-exhaustive search
over prime-power combinations, we construct $d$-disjunct matrices of
size at least 3,600. The parameters giving designs with the smallest
number of rows are given in Table~\ref{Table:EPP2007GHa}.

\begin{table}
\centering
\begin{tabular}{lrrl}
$d$ & $t$ & $n$ & prime powers \\
\hline
2 & 82 & 4077 & $\{2^3, 3^2, 5, 7,11,13,17,23\}$\\
3 & 155 &3855 & $\{2^2,3,5,7,11,13,17,19,23,29,37\}$\\
4 & 237 &3631 & $\{2^3,3,5,7,11,13,17,19,23,29,31,37,43\}$\\
5 & 333 & 4023 & $\{2^3,3^2,5,7,11,13,17,19,23,29,31,37,41,43,53\}$\\
6 & 445 &4077 & $\{2^3,3^2,5,7,11,13,17,19,23,29,31,37,41,43,53,61\}$\\
\hline
\end{tabular}
\caption{Prime powers used for the construction of $d$-disjunct
  $t\times n$.}\label{Table:EPP2007GHa}
\end{table}

\subsubsection{Designs based on error-correcting codes}

Excellent practical superimposed codes can be constructed based on
error-correcting codes. Here, we discuss a number of techniques and
constructions we used to generate the best superimposed codes known to
us for a variety of pixel array sizes (i.e, for square $m\times m$
arrays with $m\in\{10,20,30,40,60,120\}$). For comparison with other
constructions, the results are summarized in
Table~\ref{Table:ECCBasedCodes}, and in row (n) of
Table~\ref{Table:BoundsOnTnd} for $n=3,\!600$.

\paragraph{Binary codes.} The most straightforward way of obtaining
$d$-disjunct superimposed codes is by simply taking constant-weight
binary error-correction codes obeying
\eqref{Eq:LowerBoundDisjunctness}, with overlap $\mu$ as given
below. The on-line repository \cite{BROxxxxa} lists the best known
lower bound on maximum size $A(n,d,w)$ for constant weight $w$ codes
of length $n$ and Hamming distance $d$ (note the different use of $n$
and $d$ in this context). Given a code \stb{n,d,w} from this Standard
Table, the overlap satisfies $\mu \leq w - d/2$. Some codes are given
explicitly, whereas others require some more work to instantiate. We
discuss two of the most common constructions that are used to
instantiate all but one of the remaining codes we use. The first
construction consists of a short list of seed codewords $v_i$, along
with a number of cyclic permutations. These permutations give rise to
a permutation group $\mathcal{P}$ and the words in the final code are
those in the union of orbits of the seed vectors: $\cup_i \{w \mid w =
P(v_i), P\in\mathcal{P}\}$. The second construction shortens existing
codes in one of two ways. We illustrate them based on the generation
of the $2$-disjunct code \stb{21,8,7} of size 100 from
\stb{24,8,8}. The first type of shortening consists of identifying a
row $i$ with the most ones and then selecting all columns incident to
this row and deleting row $i$. This both reduces the weight and the
length of the code by one, but preserves the distance, and in this
case gives \stb{23,8,7} (note that shortenings do not in general lead
to new optimal or best known codes). The second type of shortening
identifies a row $i$ with the most zero entries and creates a new code
by keeping only the columns with a zero in the $i$-th row, and then
removing the row. This construction does not affect the weight of the
matrix, only the size and length. Repeating this twice from
\stb{23,8,7} yields the desired \stb{21,8,7} code. Note that for
constant-weight codes, the minimum number of ones in any given row is
bounded above by $\lfloor wn/t\rfloor$. This expression can be used to
give a theoretical minimum on the number of rows by which we can
shorten a code. In practice it may be possible to remove a
substantially larger number of rows. Below we will frequently use
shortening to obtain codes with smaller length. In these cases we only
use the second type of shortening, based on zero entries.

\paragraph{q-ary error-correction codes.} The Standard Table only
lists codes with relatively small lengths and consequently, limited
sizes. In order to construct larger or heavier codes we apply a
construction based on (maximal distance separable (MDS)) $q$-ary
error-correcting codes, as proposed by Kautz and Singleton
\cite{KAU1964Sa}. Let $(n,k,d)_q$ denote a linear $q$-ary code of
length $n$, size $q^k$, and Hamming distance $d$. Each codeword in
these codes consists of $n$ elements taken from $\mathrm{GF}(q)$ and
differs in at least $d$ locations from all other codewords. A binary
superimposed code can be obtained from these $q$-ary codes by
replacing each element with a corresponding column of a $q\times q$
identity matrix $I_q$. That is, we map each element in
$\mathrm{GF}(q)$ to a unique column of $I_q$.  For example, we map
value $k$ to the $(k+1)$-st column of $I_q$ as follows:
\begin{center}
\begin{tabular}{lp{6pt}cccccccc}
$q$-ary & & & 2 &0 &1 &2 &1 &1 &0 \\[6pt]
&& {\tiny 0} & 0& 1 & 0 & 0 & 0 & 0 & 1 \\
binary && {\tiny 1} & 0 & 0 & 1 & 0 & 1 & 1 & 0\\
&&{\tiny 2} & 1 & 0 & 0 & 1 & 0 & 0 & 0
\end{tabular}
\end{center}

The overlap between any two codewords of the resulting concatenated
code is bounded by the length of the code $n$ minus the distance
$d$. Meanwhile, the weight is exactly the length. The disjunctness of
the resulting code is therefore at least $\lceil (n-1) /
(n-d)\rceil$. As an aside, note that this construction requires an
explicit set of codewords. This can be contrasted with $q$-ary
error-correction codes, for which fast encoding-decoding algorithms
often exist, and which do not require such an explicit representation.

Existence of a large class of $(n,k,d)_q$ codes, including the
well-known Reed-Solomon codes is shown by Singleton
\cite{KAU1964Sa,SIN1964a}. In MacWilliams and Sloane \cite[Ch. 11,
Thm. 9]{MAC1981Sa} cyclic MDS codes with $n=q+1$ and $d=q-k+2$ are
shown to exist whenever $1 \leq k \leq q+1$ and $q$ is a prime power.
As a result, it can be concluded that, when expressed in group-testing
notation, this concatenated code construction requires a length $t =
\mathcal{O}(\min[n,k^2\log^2 n])$ \cite{POR2008Ra}, compared to the
best-known bound of $\mathcal{O}(k^2\log n)$. Despite the slightly
weaker bound, we shall see below that for small instances, the
resulting codes are far superior to the random constructions used to
yield the improved bound.

As an example, we used a concatenation of $(10,4,7)_{11}$ with
$I_{11}$, denoted $(10,4,7)_{11}^{I_q}$, to construct the 3-disjunct
matrix of size $n=14,\!400$ shown in Table~\ref{Table:ECCBasedCodes}.
Most of the other codes obtained using this construction have an
additional superscript $s(k)$ to indicate the application of $k$
shortening steps. The 4-disjunct code of length $n=1,\!600$ also has a
superscript $x$ to indicate extension of the code. In this case, the
four shortening steps resulted in a $113\times 1,\!596$ code, falling
just short of the desired size of $1,\!600$. The very structured
nature of these concatenated codes means that many constant-weight
vectors are not included, even if they may be feasible. We can
therefore try to apply greedy search techniques to augment the
code. For this particular case it was found to be relatively easy to
add several columns, thus resulting in a code of desired size.

All of the $q$-ary codes appearing in Table~\ref{Table:ECCBasedCodes}
are Reed-Solomon codes, except for $(10,4,7)_9$, which is a
constacyclic linear code whose construction is given by
\cite{GRAxxxxa}.

\paragraph{General concatenated codes.}

As mentioned by Kautz and Singleton \cite{KAU1964Sa}, it is possible
to replace the trivial identity code by arbitrary $d$-disjunct binary
matrices in forming concatenated codes, provided that the size of the
matrix is at least $q$. This construction is extensively used by
D'yachkov {\it et al.}\ \cite{DYA2000MRa} to form previously unknown
instances of superimposed codes. We also investigated this approach
and found that the concatenation of the $(7,4,4)_{11}$ code with
\stb{9,4,3} yielded our smallest $2$-disjunct code of size
14,400. Likewise a $2$-disjunct $51\times 4,\!489$ matrix can be
obtained by concatenating $(3,2,2)_{67}$ with \stb{17,6,5}. The code
given in Table~\ref{Table:ECCBasedCodes} lists instead \stb{51,8,7},
which yields a slightly smaller code with weight 7 instead of 15. Note
that concatenation with non-trivial disjunct matrices can result in
codes in $D_w(d,n) \setminus D_{w,\mu}(d,n)$ or even $D(d,n) \setminus
D_w(d,n)$.

\paragraph{Other constructions.}

We would like to mention two other construction techniques that can be
used to generate superimposed codes. The first technique is based on
certain designs such as $t$-designs and Steiner systems
\cite{COL2007Da}, which were used for example by Balding {\it et al.}
\cite{BAL1996BKTa} to construct a $4$-disjunct $65\times 520$ binary
code. The matrix corresponding to this particular system can be found
on-line in the La Jolla Covering Repository \cite{GORxxxxa} as
$C(65,9,3)$. In Table~\ref{Table:ECCBasedCodes} we shorten this matrix
by a single row to obtain a code for $n=400$. Although this
construction is significantly shorter than the best $q$-ary based
design we could find (a $76\times 408$ shortening of $(9,3,7)_9$), it
is typically much harder to obtain explicit instantiations of designs,
or even to show their existence, at least compared to $q$-ary codes.
The second technique based on Latin squares appears in
\cite{KAU1964Sa}, and was used by D'yachkov {\it et al.}
\cite{DYA2000MRa} to create $2$-disjunct $51\times 4,\!624$ and
$63\times 14,\!400$ matrices.

\begin{table}
\centering
\begin{tabular}{|r|ll|ll|ll|}
\multicolumn{1}{c}{} & \multicolumn{2}{c}{$n=100$} &
\multicolumn{2}{c}{$n=400$} & \multicolumn{2}{c}{$n=900$}  \\
\hline
\multicolumn{1}{|c|}{$d$} & $t$ & construction & $t$ & construction &
$t$ & construction \\
\hline
$2$ &
21 & \stb{21,8,7} &
31 & \stb{31,8,7} &
38 & \stb{38,8,7} \\
$3$ &
36 & \stb{36,10,7} &
51 &\stb{51,10,7} &
73 & (7,3,5)$_{11}^{I_q,s(4)}$ \\
$4$ &
48 & \stb{48,8,5} &
64 &\ctb{65,9,3}$^{s(1)}$ &
95 &(9,3,7)$_{11}^{I_q,s(4)}$ \\
$5$ &
60 & \stb{60,10,6} &
107 & (11,3,9)$_{11}^{I_q,s(14)}$ &
117 & (11,3,9)$_{11}^{I_q,s(4)}$ \\
$6$ &
75 & (7,2,6)$_{11}^{I_q,s(2)}$ &
144 & (13,3,11)$_{13}^{I_q,s(25)}$ &
156 & (13,3,11)$_{13}^{I_q,s(13)}$ \\
\hline
\multicolumn{7}{c}{}\\
\multicolumn{1}{c}{} & \multicolumn{2}{c}{$n=1,\!600$} &
\multicolumn{2}{c}{$n=3,\!600$} & \multicolumn{2}{c}{$n=14,\!400$}  \\
\hline
\multicolumn{1}{|c|}{$d$} & $t$ & construction & $t$ & construction &
$t$ & construction \\
\hline
$2$ &
44 & \stb{44,8,7} &
51 & \stb{51,8,7} &
63 & (7,4,4)$_{11}^{\stb{9,4,3}}$ \\
$3$ &
78 & (10,4,7)$_9^{I_q,s(12)}$ &
85 & (10,4,7)$_{9}^{I_q,s(5)}$ &
110 & (10,4,7)$_{11}^{I_q}$ \\
$4$ &
113 & (9,3,7)$_{13}^{I_q,s(4),x}$ &
142 & (9,3,7)$_{16}^{I_q,s(2)}$ &
161 & (13,4,10)$_{13}^{I_q,s(8)}$ \\
$5$ &
140 & (11,3,9)$_{13}^{I_q,s(3)}$ &
174 & (11,3,9)$_{16}^{I_q,s(2)}$ &
233 & (16,4,13)$_{16}^{I_q,s(23)}$\\
$6$ &
166 & (13,3,11)$_{13}^{I_q,s(3)}$ &
206 & (13,3,11)$_{16}^{I_q,s(2)}$ &
323 & (13,3,11)$_{25}^{I_q,s(2)}$\\
\hline
\end{tabular}
\caption{Overview of best-known constructions for $d$-disjunct
  matrices of size $n$. Codes from the Standard Table \cite{BROxxxxa}
  and covering are indicated by $\stb{\cdot}$ and $\ctb{\cdot}$; all
  other entries are $(n,k,d)_q$ codes. Superscripts $s(k)$ indicates a
  shortening of $k$ steps with pivoting on zero entries; $x$ indicates
  a greedy extension, $I_q$ indicates concatenation with a $q\times q$
  identity matrix. The $\stb{\cdot}$ superscript indicates concatenation
  with a binary code from the Standard Table.}\label{Table:ECCBasedCodes}
\end{table}

\begin{figure}
\centering
\includegraphics[width=0.6\columnwidth]{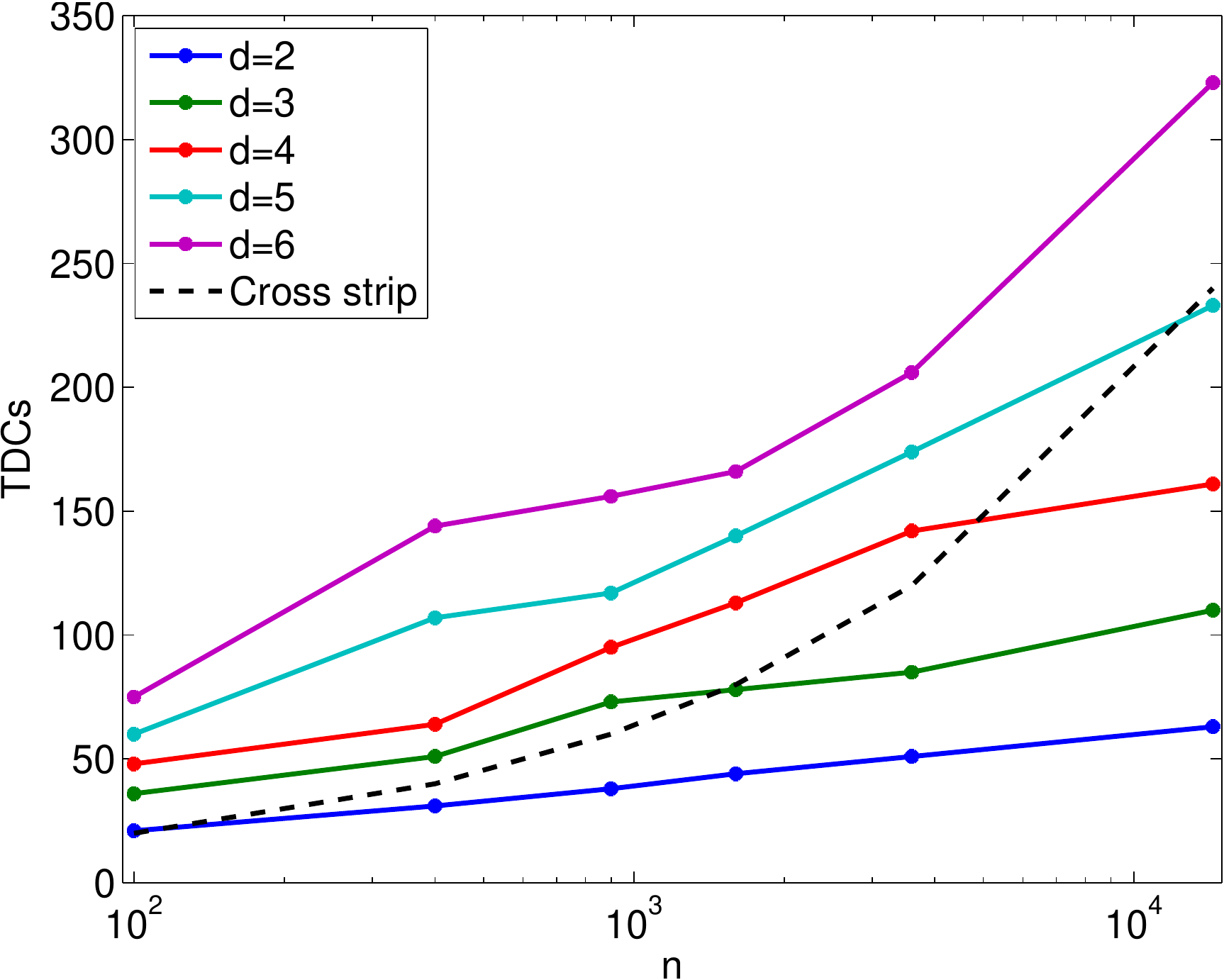}
\caption{Comparison between the number of TDCs used for the best-known
  codes in Table~\ref{Table:ECCBasedCodes} and a standard cross-strip
  design, for different numbers of pixels $n$. The cross-strip
  design connects each row and column to a single TDC, giving a $d=1$
  disjunct encoding. The number of TDCs in the cross-strip design
  grows as $\mathcal{O}(\sqrt{n})$, much faster than the
  $\mathcal{O}(d^2\log{n})$ in group-testing matrices.}
\end{figure}

\section{Theoretical bounds}

We now summarize an extensive literature concerning theoretical bounds
on the minimum number of pools required for a group-testing design of
size $n$ to be $d$-disjunct. These results include both upper and
lower bounds on this minimum, and can be used to get an idea about the
quality of the matrices we constructed in the previous section.

\subsection{Asymptotic bounds}\label{Sec:AsymptoticBounds}

Bounds on the growth rate of $T_D(n,d)$ have been discovered and
rediscovered in different contexts in information theory,
combinatorics, and group testing~\cite{RUS1994a}. In the context of
superimposed codes, D'yachkov and Rykov~\cite{DYA1983Ra} obtained the
following bounds:
\[
\Omega(d^2\log n / \log d) \leq T_D(n,d) \leq \mathcal{O}(d^2\log n).
\]
Ruszink\'{o}~\cite{RUS1994a} and F\"{u}redi~\cite{FUR1996a} give an
interesting account on the lower bound and provide simpler proofs. The
lower bound was also obtained for sufficiently large $d$ with $d\leq
n^{1/3}$ by Chaudhuri and Radha\-krishnan~\cite{CHA1996Ra} in the
context of $d$-cover free systems for system complexity, which was
extended to the general case by Clementi {\it et
  al.}~\cite{CLE2003MSa}.

For the upper bound, it follows from the analysis of a greedy-type
construction given by Hwang and S\'{o}s~\cite{HWA1987Sa}, that
for $t \geq 16d^2$ we have
\begin{equation}\label{Eq:BoundHWA1987Sa}
T_{D_{w,\mu}}(n,d) \leq 16 d^2\cdot \log_3(2)\cdot(\log_2(n)-1) < 11d^2\log_2(n).
\end{equation}
An efficient polynomial-time algorithm for generating similar
constant-weight $d$-disjunct $t\times n$ matrices with $t =
\mathcal{O}(d^2\log n)$ was given by Porat and Rothschild
\cite{POR2008Ra}.

\subsection{Non-asymptotic results}

When working with particular values of $d$ and $n$, constants in
theoretic bounds become crucial. Most theoretical results, however,
are concerned with growth rates and even if explicit constants are
given, such as those in \eqref{Eq:BoundHWA1987Sa}, they may be too
loose to be of practical use.

In this section we are interested in the code length $t$ required to
guarantee the unique recovery of up to $d=6$ simultaneous firings on a
$60\times 60$ pixel array, that is $n=$3,600. We do this by studying
the various models used to obtain upper and lower bounds on $T(n,d)$
and numerically evaluating the original expressions, instead of
bounding them. By optimizing over parameters such as weight or matrix
size, we obtain the best numerical values for the bounds, as permitted
by the different models. These will be used as a reference point to
evaluate the quality of the codes given in
Table~\ref{Table:ECCBasedCodes}.

All bounds are evaluated using Sage \cite{Software:SAGE} and
summarized in Table~\ref{Table:BoundsOnTnd} for $d=2,\ldots,6$. We
omit results for $d=1$, which only requires the columns of the matrix
to be unique.

\subsubsection{Upper bounds on $T(n,d)$}

The construction of a $d$-disjunct binary $t\times n$ matrix, or
providing an algorithm for doing so, immediately yields $t$ as an
upper bound for one of the $T(n,d)$ values. In the literature there
are three main techniques to obtain $d$-disjunct matrices of various
sizes; we discuss each of these in the following paragraphs.

\paragraph{Sequential picking.} The construction given by Hwang and
S\'{o}s \cite{HWA1987Sa} for obtaining an upper bound on
$T_{D_{w,\mu}}(n,d)$ works as follows. Starting with a family of
initially all good sets $G_0 = {[t] \choose w}$ of a fixed weight $w$
we pick a random element $g \in G_0$. After we make this choice we
determine the family $B_1$ of sets in $G_0$ whose overlap with $g$ is
$w/d$ or more (this includes $g$ itself). We then remove all the bad
items $B_1$ from $G_0$ to get $G_1 = G_0 \setminus B_1$. We then
repeat the procedure, picking at each point a $g$ from $G_i$,
determining $B_{i+1}$ and forming $G_{i+1}$, until $G_n$ is empty. We
can then form a $t \times n$ matrix $A$ with the support of each
column corresponding to one of the selected $g$. This matrix will be
$d$-disjunct since, by construction, it has constant column weight $w$
and satisfies \eqref{Eq:LowerBoundDisjunctness} with overlap $\mu <
w/d$. For the number of rows in $A$, notice that the size of each
$B_i$ satisfies
\[
\vert B_i\vert \leq \sum_{i=\lceil w/d\rceil}^w {w\choose
  i}{t-w\choose w-i}
\]
Because the size of the initial family $G_0$ was ${t\choose w}$ we
have that
\begin{equation}\label{Eq:BoundSequentialPicking}
n \geq {t\choose w}\left/ \sum_{i=\lceil w/d\rceil}^w {w\choose
  i}{t-w\choose w-i}\right..
\end{equation}
This can be turned into the upper bound on $T_{D_{w,\mu}}(n,d)$ shown
in \eqref{Eq:BoundHWA1987Sa} by following the parameter choice and
analysis given in \cite{HWA1987Sa}. Essentially the same argument was
used earlier by Erd\"{o}s {\it et al.}\ to prove an alternative bound
given in \cite[Proposition 2.1]{ERD1985FFa}. Entries (a--c) in
Table~\ref{Table:BoundsOnTnd} are obtained using respectively
\eqref{Eq:BoundHWA1987Sa}, the bound given in \cite{ERD1985FFa} with
optimal choice of $w$, and the smallest $t$ that satisfies
\eqref{Eq:BoundSequentialPicking} for some $w \leq t$ (see also
\cite[p.~232]{DEB2003Va}).

\paragraph{Random ensemble.} An alternative to picking compatible
columns one at a time, is to draw an entire set of columns from a
suitable distribution and check if the resulting matrix is indeed
$d$-disjunct. An upper bound on the required number of rows is
obtained by finding the smallest $t$ such that the probability that
the matrix does not satisfy $d$-disjunctness is strictly bounded above
by one.

Consider a $t\times n$ matrix $A$ with entries drawn i.i.d.\ from the
Bernoulli distribution; each entry takes on the value 1 with
probability $\beta$ and the value $0$ with probability $1-\beta$. For
any column in $A$, the probability that it is covered by $d < n$ of
the other columns is given by
\begin{equation}\label{Eq:ProbBernoulliCovered}
p = (1 - \beta(1-\beta)^d)^t,
\end{equation}
since each entry in the column is not covered by the union of the
other $d$ if and only if its value is 1, and the corresponding entries
in the other $d$ columns are all 0. Since we want to minimize the
chance of overlap, we find the minimum with respect to $\beta$, giving
$\beta = 1/(d+1)$. By taking a union bound over all possible sets, we
can bound the probability that at least one column in $A$ is covered
by some $d$ others as
\begin{equation}\label{Eq:UnionBoundGroups}
P(A\ \mathrm{not}\ d\mathrm{-disjunct}) \leq n{n-1\choose d} p.
\end{equation}
This setup forms the basis for the analysis given by D'yachkov and
Rykov \cite{DYA1983Ra}. In Table~\ref{Table:BoundsOnTnd}, row (d), we
show the smallest $t$ for which this probability is less than one;
for the optimal $\beta$, this is given by
\[
t = \left\lfloor 1 - \log\left(n{n-1 \choose d}\right) \left/ \log\left( 1
      - \frac{d^d}{(d+1)^{d+1}}\right)\right.\right\rfloor.
\]

To construct matrices with constant-weight $w$, we can uniformly draw
columns from $S = {[t] \choose w}$. The probability that a fixed
column is covered by $d$ others can be evaluated as follows. Let
$R_d(t,w,i)$ denote the number of ways we can cover a fixed set of $w$
entries in a $t$-length vector with $d$ columns, given that $i$
entries have already been covered. With only a single column left, we
first cover the remaining $w-i$ entries, leaving $i$ entries to be
freely selected in the remaining $t-(w-i)$ positions. This gives
\[
R_1(t,w,i) = {t-w+i \choose i}.
\]
For $d>1$ we can cover up to $w-i$ entries and cover the others
using the remaining $d-1$ columns. Therefore,
\[
R_{d}(t,w,i) = \sum_{j=0}^{w-i}{w-i \choose j}{t-w+j \choose w-j}R_{d-1}(t,w,i+j).
\]
The probability that a column is covered by $d$ others is then given
by
\begin{equation}\label{Eq:ProbConstWeightCovered}
p = R_d(t,w,0) \Big/  {t \choose w}^{d},
\end{equation}
and we can again use the union bound given by
\eqref{Eq:UnionBoundGroups}. In order to determine the smallest
possible $t$ we start at $t=1$ and double its value until the
right-hand side of \eqref{Eq:UnionBoundGroups} is less than one and
then use binary search to find the desired $t$. This is repeated for
all suitable column weights $w$ and row (e) of
Table~\ref{Table:BoundsOnTnd} lists the smallest value of $t$ obtained
using this procedure.

The use of the union bound in \eqref{Eq:UnionBoundGroups} ignores the
fact that many of the columns and $d$-sets are independent. In order
to obtain sharper bounds for a similar problem, we follow Yeh
\cite{YEH2002a} and apply Lov\'{a}sz Local Lemma \cite{ERD1974La},
stated in \cite[Corollary 5.1.2]{ALO1992Sa}.

\begin{lemma}[The Local Lemma; Symmetric case]\label{Lemma:LocalLemma}
  Let $E_1$, $E_2$,\ldots,$E_n$ be events in an arbitrary probability
  space. Suppose that each event $E_i$ is independent of all other
  events $E_j$ except at most $\mu$ of them, and that $P(E_i)
  \leq p$ for all $1\leq i\leq n$. If
\begin{equation}\label{Eq:ConditionLocalLemma}
ep(\mu+1) \leq 1
\end{equation}
then $P(\cap_{i=1}^n\bar{E}_i) > 0$.
\end{lemma}

Working again with $n$ columns drawn uniformly and independently from
${[t]\choose n}$, let $E_{i,J}$ denote the event that column $i$ is
covered by the set $J$ of $d$ other columns. Any event $E_{i_1,J_1}$
is independent from $E_{i_2,J_2}$ whenever $(J_1 \cup \{i_1\}) \cap
(J_2 \cup \{i_2\}) = \emptyset$, and the number of events $\mu$ that
violate this condition is given by
\[
\mu = (d+1)\left[{n \choose d+1} - {n-d-1 \choose d+1} \right].
\]
All we then need to do is find the smallest $t$ for which, with an
optimal choice of $w$, condition \eqref{Eq:ConditionLocalLemma} is
satisfied using $p$ as given in
\eqref{Eq:ProbConstWeightCovered}. That is, find a $t$ such that
\[
e\cdot R_d(t,w,0){t \choose w}^{-d}(\mu + 1) \leq 1.
\]
The results obtained by this method are given in row (f) in
Table~\ref{Table:BoundsOnTnd}. We omit similar results obtained using
a Bernoulli model, or the $q$-ary construction considered by
Yeh~\cite{YEH2002a}, since they give larger values of $t$.

\paragraph{Random selection with post processing.} A third approach to
generating $d$-disjunct matrices is to start with a $t\times m$ matrix
with $m\geq n$, drawn from a certain random ensemble, and randomly
mark for deletion one of each pair of columns whose overlap is too
large. Whenever the expected number of columns marked for deletion is
strictly less than $n-m+1$ we can be sure that a $d$-disjunct $t\times
n$ matrix exists and can be generated using this approach. (Note that
some of the approaches described in the previous paragraph are a
special case of this with $m=n$.) A variation on this approach is to
pick a column and mark it for deletion if there exist any other $d$
columns that cover it. Starting with this latter approach for the
Bernoulli model with $\beta = 1/(d+1)$, a bound on the probability of
a column being marked is found by applying a union bound to
\eqref{Eq:ProbBernoulliCovered}:
\[
P(\mathrm{covered}) \leq {m-1 \choose d}\left(1 - \frac{d^d}{(d+1)^{d+1}}\right)^t.
\]
The expected number of columns marked is then bounded by
\[
\mathbb{E}(\#\mathrm{cols}) = m\cdot P(\mathrm{covered})
\leq m {m-1 \choose d}\left(1 - \frac{d^d}{(d+1)^{d+1}}\right)^t,
\]
and the smallest $t$ for which the left-hand side of this inequality
is strictly less than $m-n+1$ is given in row (g) of
Table~\ref{Table:BoundsOnTnd}. A similar derivation based on
constraints on the pairwise overlap of columns can easily be seen to
yield
\[
\mathbb{E}(\#\mathrm{columns\ marked}) \leq {m\choose 2}\cdot{t\choose
  w}^{-1}\sum_{i=\lceil w/d\rceil}^{w}{w\choose i}{t-w \choose w-i},
\]
which leads to the results shown in row (h).  The best results for
$T_{D}(n,d)$ based on bounds of the probabilistic method in row (i)
were obtained for groupwise covering of constant weight vectors, a
model which was studied earlier by Riccio and Colbourn
\cite{RIC1998Ca}. For completeness we list in row (j) the results
based on a random $q$-ary construction similar to the one we discussed
earlier. A final method we would
like to mention chooses $m=n$, but instead of removing marked columns,
it augments the matrix with the identity matrix below all marked
columns, thus fixing a (potential) violation of $d$-disjunctness by
increasing the number of rows. The results of this method by
Yeh~\cite{YEH2002a} are presented in row (k).

\begin{table}[t]
\centering
\begin{tabular}{|cr|rrrrr|l|}
\hline
& \multicolumn{1}{l}{Bound\hfill$\setminus$\hfill$d$} & 2 & 3 & 4 & 5 & 6 & Reference and comments \\
\hline\hline
\multicolumn{8}{|c|}{ Upper bounds (theoretical)}\\
\hline
(a) & $T_{D_{w,\mu}}(n,d) \leq$ & 436 & 982 & 1746 & 2729 & 3929 &
Adaptation of \cite{HWA1987Sa} given in \eqref{Eq:BoundHWA1987Sa}\\
(b) & $T_{D_{w,\mu}}(n,d) \leq$  & 94 & 237 & 443 & 711 & 1043 &
\cite[Proposition 2.1]{ERD1985FFa} with optimal $w$ \\
(c) & $T_{D_{w,\mu}}(n,d) \leq$ & {\bf 90} & {\bf 222} & {\bf 412} &
{\bf 660} & {\bf 966} &
Equation \eqref{Eq:BoundSequentialPicking} with optimal $w$ \\
\hline
(d) & $T_{D}(n,d) \leq $ & 149 & 278 & 442 & 640 & 870 & Bernoulli, 
\cite[Theorem 5]{DYA1983Ra} \\
(e) & $T_{D_w}(n,d) \leq$ & 96 & 190 & 312 & 459&  631 & Constant
weight, groupwise cover \\
(f) & $T_{D_w}(n,d) \leq$ & 76 & 163 & 279 & 422 & 590 & Constant
weight, Lov\'{a}sz local lemma\\
\hline
(g) &$T_{D}(n,w) \leq$ & 110 & 225 & 376 & 561 & 779 &
Bernoulli, groupwise cover \\
(h) & $T_{D_{w,\mu}}(n,d) \leq$ & 99 & 249 & 465 & 746 & 1094 &
Constant weight, pairwise overlap\\
(i) & $T_{D_{w}}(n,d) \leq$ & {\bf 71} & {\bf 154} & {\bf 265} & {\bf
  402} & {\bf 565} &
Constant weight, groupwise cover; see \cite{RIC1998Ca}\\
(j) & $T_{D_{w,\mu}}(n,d) \leq$ & 130 & 300 & 522 & 828 & 1178 &
Adaptation of \cite[Algorithm 1]{CHE2008Da}\\
(k) & $T_{D}(n,d) \leq$ & 98 & 205 & 348 & 524 & 734 &
\cite[Theorem 2.6]{YEH2002a}\\
\hline
\multicolumn{8}{|c|}{Upper bounds (instances)}\\
\hline
(l) & $T_{D_{w,\mu}}(n,d) \leq$ & 58 & 132 & 224 & 345 & 484 &
Greedy search\\ 
(m) &  $T_{D_w}(n,d) \leq$ & 82 & 155 & 237 & 333 & 445 &
\cite[Section
2]{EPP2007GHa}; see Table~\ref{Table:EPP2007GHa}\\
(n) & $T_{D_{w,\mu}}(n,d) \leq$  & 51 & 85 & 142 & 174 & 206
&
Error-correction code based\\ 
\hline
\multicolumn{8}{|c|}{Lower bounds}\\
\hline
(o) & $T_{D_{w,\mu}}(n,d) \geq$ & {\bf 35} & {\bf 56} & {\bf 75} &
{\bf 96} & {\bf 115} & \cite[Theorem 2]{JOH1963a} \\
(p) & $T_{D_{w,\mu}}(n,d) \geq$ & 35 & 55 & 74 & 93 & 113 &
\cite[Equation 5]{KAU1964Sa} and \cite[Equation 18]{DYA1983Ra}\\
(q) & $T_{D_w}(n,d) \geq$ & {\bf 35} & -- & -- & -- & -- & \cite[Theorem 1]{ERD1982FFa}\\
(r) & $T_{D_{w,\mu}}(n,d) \geq$ & 32 & 48 & 62 & 75 & 88 & \cite[Proposition 2.1]{ERD1985FFa} \\
(s) & $T_{D_w}(n,d) \geq$ & 29 & {\bf 40} & {\bf 48} & {\bf 55} & 61 & \cite[Lemma
3.2]{RUS1994a} \\
(t) & $T_S(n,d) \geq$ & {\bf 23} & {\bf 33} & {\bf 43} & {\bf 53} &
{\bf 62} &
Information theoretical lower bound \\
\hline
\end{tabular}
\caption{Upper and lower bounds on the minimum number of rows needed
  for a $d$-disjunct or $\bar{d}$-separable matrix with $n=3600$
  rows. Entries shown in boldface denote the best theoretical upper
  and lower bounds we could find for each of the given classes.}\label{Table:BoundsOnTnd}
\end{table}

\paragraph{Remarks on the upper bounds.}

Even though different models may lead to the same growth rates in
$T(n,d)$, there is a substantial difference in resulting values, when
numerically evaluated. The groupwise models easily outperform bounds
based on pairwise comparison overlap. In addition there is a large gap
between matrices with constant-weight columns, and those generated
with i.i.d.\ Bernoulli entries, the former giving much more favorable
results. The second approach given above, based on drawing fixed
$t\times n$ matrices is clearly outperformed by the third approach in
which a $t\times m$ matrix is screened and reduced in size by removing
columns that violate disjunctness or maximum overlap conditions, even
if the Lov\'{a}sz Local Lemma is used to sharpen the bounds.

\subsection{Lower bounds on {\small$\mathbf{T(n,d)}$}}

Most of the lower bounds listed in Table~\ref{Table:BoundsOnTnd} are
derived using the concept of a \emph{private set}. Let $\mathcal{F}$
be a family of sets (the columns of our matrix $A$). Then $T$ is a
private set of $F \in\mathcal{F}$, if $T \subseteq F$, and $T$ is not
included in any of the other sets in $\mathcal{F}$. For $\mathcal{F}
\in D_{w,\mu}(n,d)$ with length $t$, constant column weight $w$, and
maximum overlap $\mu$, it is easily seen that, by definition, each
$(\mu+1)$-subset of $F \in\mathcal{F}$ is private. The number of such
private sets $\vert\mathcal{F}\vert\cdot {w \choose \mu+1}$ cannot
exceed the total number of these sets, ${t \choose \mu+1}$, and it
therefore follows \cite{DYA1983Ra,KAU1964Sa} that
\[
\vert \mathcal{F}\vert  \leq  {t \choose \mu+1}\big/ {w \choose \mu+1}.
\]
Based on this we can find the smallest $t$ for which there are values
$w$ and $\mu$ obeying $(w-1)/\mu \geq d$, such that the right-hand
side of the above inequality exceeds the desired matrix size. This
gives a lower bound on the required code length and the resulting
values are listed in row (p) of Table~\ref{Table:BoundsOnTnd}. A
slightly better bound on the same class of matrices follows from an
elegant argument due to Johnson~\cite{JOH1963a}, and is given in row
(o).

Ruszink\'{o} \cite{RUS1994a} studies $d$-disjunctness in
constant-weight matrices without considering maximum overlap and
provides the following argument. Let $\mathcal{F} \in D_{w}(n,d)$ be a
family of $w$-subsets in $[t]$, such that no $d$ sets in $\mathcal{F}$
cover any other set. Moreover, assume that $w = kd$ for some integer
$k$. Any $F \in\mathcal{F}$ can be partitioned into $d$ sets of length
$k$, and it follows from the $d$-disjunctness of $\mathcal{F}$ that at
least one of these $d$ subsets is private to $F$ (otherwise it would
be possible to cover $F$ with $d$ other columns).  The key observation
then, is that Baranyai's Theorem~\cite{BAR1973a} guarantees the
existence of $s = {w \choose w/d}/d$ different partitions of $F$ such
that no subset in the partitions is repeated. Each of these $s$
partitions contains a private set, so the total number of private sets
is at least $\vert \mathcal{F}\vert\cdot s$. Rewriting this gives
\[
\vert \mathcal{F} \vert \leq d {t \choose w/d}\Big/{w \choose w/d}.
\]
With proper rounding this can be extended to general weights $w$,
giving the results shown in row (s).
The results in rows (q) and (r) are also obtained based on private
sets, but we will omit the exact arguments used here. Finally, an
evaluation of the information-theoretic bound
\eqref{Eq:InfThLowerBound} is given in row (t).

\section{Numerical Experiments}

To illustrate the strength of the proposed design, we analyze the
performance of some of the group-testing designs given in
Table~\ref{Table:ECCBasedCodes}, when used to decode scintillation
events in a PET setting.

\subsection{Simulation parameters}

In PET, scintillation crystals are used to convert 511 kilo-electron
volt annihilation photons into bursts of low-energy (visible light)
photons. We use the standard software package
Geant4~\cite{AGO2003AAAa,ROD2004MOPa} to simulate this process for
some 1,000 scintillation events for a $3 \times 3 \times 10$ mm$^3$
cerium-doped lutetium oxyorthosilicate (LSO) crystal (7.4 g/cm$^3$,
$n=1.82$, absorption length = 50 m, scintillation yield = 26,000
photons/MeV, fast time constant = 40 ns, yield ratio = 1, resolution
scale = 4.41) coupled to a $3 \times 3 \times 0.75$ mm silicon sensor
by 50$\mu$ of optical grease ($n=1.5$, absorption length = 50m). For
each event the simulation yields the location and arrival time of the
low-energy photons with respect to the start of the event.

The silicon sensor is assumed to have a 70\% fill factor and a quantum
efficiency of 50\% for blue photons. Once a photon is detected a pixel
will be unable to detect a new photon for a known dead time, simulated
to lie between 10 and 80 ns. We assume that each pixel takes up the
same fraction of detector area and assign photons uniformly at random
to a pixel. This increases the pixel firing rate by avoiding hitting
dead pixels in high flux regions, thereby making the decoding process
more challenging. We do not model dark counts and cross-talk between
pixels, since they would account to less than 1\% of the total number
of detected photons and therefore do not significantly affect the
result.

In terms of pixel firings, we ignore the jitter in the time between an
incident photon and the actual firing of a pixel. This assumption
greatly simplifies the simulation, and is not expected to have any
significant influence on the results. The signal delays over the
interconnection network between pixels and TDCs are assumed to be
uniform, i.e., the travel time of the signal from each pixel to any
connected TDC is assumed to be the identical, and can therefore be set
to zero without loss of generality. TDCs are further assumed to be
ideal in the sense that they have no down time; a time stamp is
recorded whenever an event occurred during the sampling interval.

\subsection{Decoding}

Pixel firings give rise to specific patterns of TDC recordings through
the group-testing design embedded in the interconnection
network. These patterns can be decoded by considering the time stamps
recorded in the memory buffer associated with each TDC and forming a
binary test vector for each time interval.  Given a test vector $y$,
the decoding process starts by identifying all codewords $a_i$ that
are covered by $y$. When the group testing matrix $A$ is a
$d$-disjunct matrix, it immediately follows that whenever there are no
more than $d$ simultaneous firings, only the codewords corresponding
to those pixels will be selected, and the decoding is
successful. Whenever there are $s > d$ pixels that fired
simultaneously, many more than $s$ columns may be covered by $y$. If
none of these columns can be omitted to form $y$ then those columns
coincide with pixels that fired and decoding is again
successful. Otherwise, the decoding is ambiguous and considered
unsuccessful. When decoding is successful we recover the pixels that
fired, otherwise they are missed.

\subsection{Results}

In our simulations we considered sensors with both $60\times 60$ and
$120\times 120$ pixel arrays, and a variety of different pixel
dead times and TDC interval lengths. Due to space limitations we can
only show a select number of tables that are representative of the
results, and we will describe any significant differences in the text.

Table~\ref{Table:Decoding2b} shows the simulation results obtained
using $60\times 60$ and $120\times 120$ arrays with interconnection
networks based on the superimposed codes from
Table~\ref{Table:ECCBasedCodes}, with disjunctness ranging from 2 to
6. The pixel dead time and TDC interval length are chosen to be 20 ns
and 40 ps, respectively. These times are somewhat pessimistic; the
typical dead time is longer, while the TDC interval could be
shorter. Both choices cause an increase in the number of simultaneous
photons per TDC sampling window, thereby making recovery more
challenging. In particular, note that the maximum number of
simultaneous firings reaches 14, far above the guaranteed recovery
level. Nevertheless, it can be seen that even with a $4$-disjunct
interconnection network, more than 99\% of all pixel firings are
successfully recovered. For the $120\times 120$ array, this number is
slightly lower at 98.5\% since fewer photons hit dead pixels, thus
causing an increase in the number of firings and, consequently, an
increase in the number of simultaneous hits.

\begin{table}
\centering
\begin{tabular}{c}
\begin{tabular}{|cr||r|r|r|r|r|}
\hline
& & \multicolumn{5}{c|}{Disjunctness (\#TDCs)}\\
\cline{3-7}
&  & $d=2$ (51) & $d=3$ (85) & $d=4$ (142) & $d=5$ (174) & $d=6$ (206)\\
\hline
\#Sim. & \#Events & \multicolumn{5}{c|}{Decoded events (\%)}\\
\hline
 1 &      1,021,073 & 100.000 & 100.000 & 100.000 & 100.000 & 100.000\\
 2 &        419,174 & 100.000 & 100.000 & 100.000 & 100.000 & 100.000\\
 3 &        192,232 &  44.015 & 100.000 & 100.000 & 100.000 & 100.000\\
 4 &         87,389 &   2.289 &  89.808 & 100.000 & 100.000 & 100.000\\
 5 &         38,526 &   0.026 &  44.271 &  99.574 & 100.000 & 100.000\\
 6 &         16,126 &   0.000 &   7.590 &  96.366 &  99.895 & 100.000\\
 7 &          6,186 &   0.000 &   0.404 &  84.481 &  99.208 & 100.000\\
 8 &          2,216 &   0.000 &   0.000 &  59.206 &  94.540 &  99.549\\
 9 &            725 &   0.000 &   0.000 &  27.448 &  83.586 &  97.793\\
10 &            231 &   0.000 &   0.000 &   7.792 &  59.307 &  91.342\\
11 &             70 &   0.000 &   0.000 &   1.429 &  30.000 &  71.429\\
12 &             17 &   0.000 &   0.000 &   0.000 &   5.882 &  70.588\\
13 &              6 &   0.000 &   0.000 &   0.000 &   0.000 &   0.000\\
14 &              1 &   0.000 &   0.000 &   0.000 &   0.000 & 100.000\\
\hline
\multicolumn{2}{|r||}{\#Pixel firings} & \multicolumn{5}{c|}{Pixel firings missed (\%)}\\
\hline
&            3,145,990  &  32.571 &   9.636 &   0.833 &   0.135 &   0.025\\
\hline
\end{tabular} \\ \\[-8pt]
({\bf{a}}) $60\times 60$ pixel array \\[16pt]
\begin{tabular}{|cr||r|r|r|r|r|}
\hline
& & \multicolumn{5}{c|}{Disjunctness (\#TDCs)}\\
\cline{3-7}
&  & $d=2$ (63) & $d=3$ (110) & $d=4$ (161) & $d=5$ (233) & $d=6$ (323)\\
\hline
\#Sim. & \#Events & \multicolumn{5}{c|}{Decoded events (\%)}\\
\hline
 1 &        992,945 & 100.000 & 100.000 & 100.000 & 100.000 & 100.000\\
 2 &        444,448 & 100.000 & 100.000 & 100.000 & 100.000 & 100.000\\
 3 &        230,712 &  16.390 & 100.000 & 100.000 & 100.000 & 100.000\\
 4 &        118,381 &   0.091 &  95.275 & 100.000 & 100.000 & 100.000\\
 5 &         56,684 &   0.000 &  62.900 &  99.737 & 100.000 & 100.000\\
 6 &         25,215 &   0.000 &  15.435 &  96.764 &  99.996 & 100.000\\
 7 &         10,061 &   0.000 &   1.312 &  80.698 &  99.901 & 100.000\\
 8 &          3,711 &   0.000 &   0.054 &  43.897 &  99.111 & 100.000\\
 9 &          1,224 &   0.000 &   0.000 &  12.582 &  95.997 &  99.918\\
10 &            421 &   0.000 &   0.000 &   1.663 &  83.610 &  99.525\\
11 &            115 &   0.000 &   0.000 &   0.000 &  56.522 &  99.130\\
12 &             33 &   0.000 &   0.000 &   0.000 &  27.273 & 100.000\\
13 &              8 &   0.000 &   0.000 &   0.000 &  12.500 & 100.000\\
14 &              3 &   0.000 &   0.000 &   0.000 &   0.000 & 100.000\\
\hline
\multicolumn{2}{|r||}{\#Pixel firings} & \multicolumn{5}{c|}{Pixel firings missed (\%)}\\
\hline
&            3,599,359  &  44.554 &  10.326 &   1.430 &   0.068 &   0.001\\
\hline
\end{tabular} \\ \\[-8pt]
({\bf{b}}) $120 \times 120$ pixel array
\end{tabular}
\caption{Decoding statistics for two pixel arrays using a pixel dead
  time of 20ns and TDC interval length of 40ps. Results shown
  correspond to the decoding of 1,000 sequential scintillation
  events.}\label{Table:Decoding2b}
\end{table}

A more complete picture on the relationship between the number of
pixel firings missed and the pixel dead time and TDC interval length
is given in Table~\ref{Table:Decoding3}. Shown are the percentage of
pixel firings missed for various pixel dead times and TDC interval
lengths.
As expected, smaller TDC intervals reduce the number of simultaneous
hits and therefore lead to more uniquely decodable events.

\begin{table}
\centering
\begin{tabular}{c}
\begin{tabular}{|lrr|rrrrr|}
\hline
Dead time & \multicolumn{1}{c}{TDC} & \#Simult. & \multicolumn{5}{c|}{Pixel firings missed (\%)}\\
({\#}firings) & \multicolumn{1}{c}{interval} &  & $d\!=\!2$ & $d\!=\!3$ & $d\!=\!4$ & $d\!=\!5$ & $d\!=\!6$\\
\hline
10ns & 5ps & 6     &   1.32 &   0.02 &   0.00 &   0.00 &   0.00\\
(3.4m)  & 10ps & 7 &   4.69 &   0.25 &   0.00 &   0.00 &   0.00\\
  & 20ps & 11      &  14.48 &   1.87 &   0.04 &   0.00 &   0.00\\
  & 40ps & 14      &  35.39 &  10.52 &   0.87 &   0.13 &   0.02\\
\hline
20ns & 5ps & 6     &   1.21 &   0.02 &   0.00 &   0.00 &   0.00\\
(3.1m)  & 10ps & 7 &   4.31 &   0.23 &   0.00 &   0.00 &   0.00\\
  & 20ps & 11      &  13.30 &   1.74 &   0.04 &   0.00 &   0.00\\
  & 40ps & 14      &  32.57 &   9.63 &   0.83 &   0.13 &   0.02\\
\hline
40ns & 5ps & 6     &   1.22 &   0.03 &   0.00 &   0.00 &   0.00\\
(2.8m)  & 10ps & 7 &   4.33 &   0.25 &   0.00 &   0.00 &   0.00\\
  & 20ps & 11      &  13.21 &   1.83 &   0.05 &   0.00 &   0.00\\
  & 40ps & 14      &  31.60 &   9.95 &   0.90 &   0.14 &   0.02\\
\hline
80ns & 5ps & 6     &   1.34 &   0.03 &   0.00 &   0.00 &   0.00\\
(2.5m)  & 10ps & 7 &   4.73 &   0.28 &   0.00 &   0.00 &   0.00\\
  & 20ps & 11      &  14.33 &   2.03 &   0.05 &   0.00 &   0.00\\
  & 40ps & 14      &  33.83 &  11.04 &   1.00 &   0.16 &   0.03\\
\hline
\end{tabular} \\ \\[-8pt]
({\bf{a}}) $60\times 60$ pixel grid\\[16pt]
\begin{tabular}{|lrr|rrrrr|}
\hline
Dead time & \multicolumn{1}{c}{TDC} & \#Simult. & \multicolumn{5}{|c|}{Pixel firings missed (\%)}\\
(\#firings) & \multicolumn{1}{c}{interval} &  & $d=2$ & $d=3$ & $d=4$ & $d=5$ & $d=6$\\
\hline
10ns & 5ps & 6 &   2.27 &   0.02 &   0.00 &   0.00 &   0.00\\
(3.7m)  & 10ps & 8 &   7.58 &   0.20 &   0.00 &   0.00 &   0.00\\
  & 20ps & 12 &  21.33 &   1.69 &   0.08 &   0.00 &   0.00\\
  & 40ps & 14 &  45.52 &  10.64 &   1.46 &   0.07 &   0.00\\
\hline
20ns & 5ps & 6 &   2.21 &   0.02 &   0.00 &   0.00 &   0.00\\
(3.6m)  & 10ps & 8 &   7.39 &   0.20 &   0.00 &   0.00 &   0.00\\
  & 20ps & 12 &  20.80 &   1.64 &   0.08 &   0.00 &   0.00\\
  & 40ps & 14 &  44.55 &  10.33 &   1.43 &   0.07 &   0.00\\
\hline
40ns & 5ps & 6 &   2.20 &   0.02 &   0.00 &   0.00 &   0.00\\
(3.5m)  & 10ps & 8 &   7.35 &   0.20 &   0.00 &   0.00 &   0.00\\
  & 20ps & 12 &  20.62 &   1.66 &   0.08 &   0.00 &   0.00\\
  & 40ps & 14 &  43.95 &  10.37 &   1.46 &   0.07 &   0.00\\
\hline
80ns & 5ps & 6 &   2.25 &   0.02 &   0.00 &   0.00 &   0.00\\
(3.4m)  & 10ps & 8 &   7.50 &   0.20 &   0.00 &   0.00 &   0.00\\
  & 20ps & 12 &  20.97 &   1.71 &   0.08 &   0.00 &   0.00\\
  & 40ps & 14 &  44.46 &  10.64 &   1.50 &   0.07 &   0.00\\
\hline
\end{tabular} \\ \\[-8pt]
({\bf{b}}) $120\times 120$ pixel grid
\end{tabular}
\caption{Percentage of pixel firings missed for various combinations
  of pixel dead time, TDC interval length, and disjunctness. Results
  shown correspond to the decoding of 1,000 sequential scintillation
  events. The first three columns show the dead time (and total number
  of pixel firings), TDC time interval, and maximum number of
  simultaneous hits, respectively.}\label{Table:Decoding3}
\end{table}

\begin{figure}
\centering
\includegraphics[width=0.65\columnwidth]{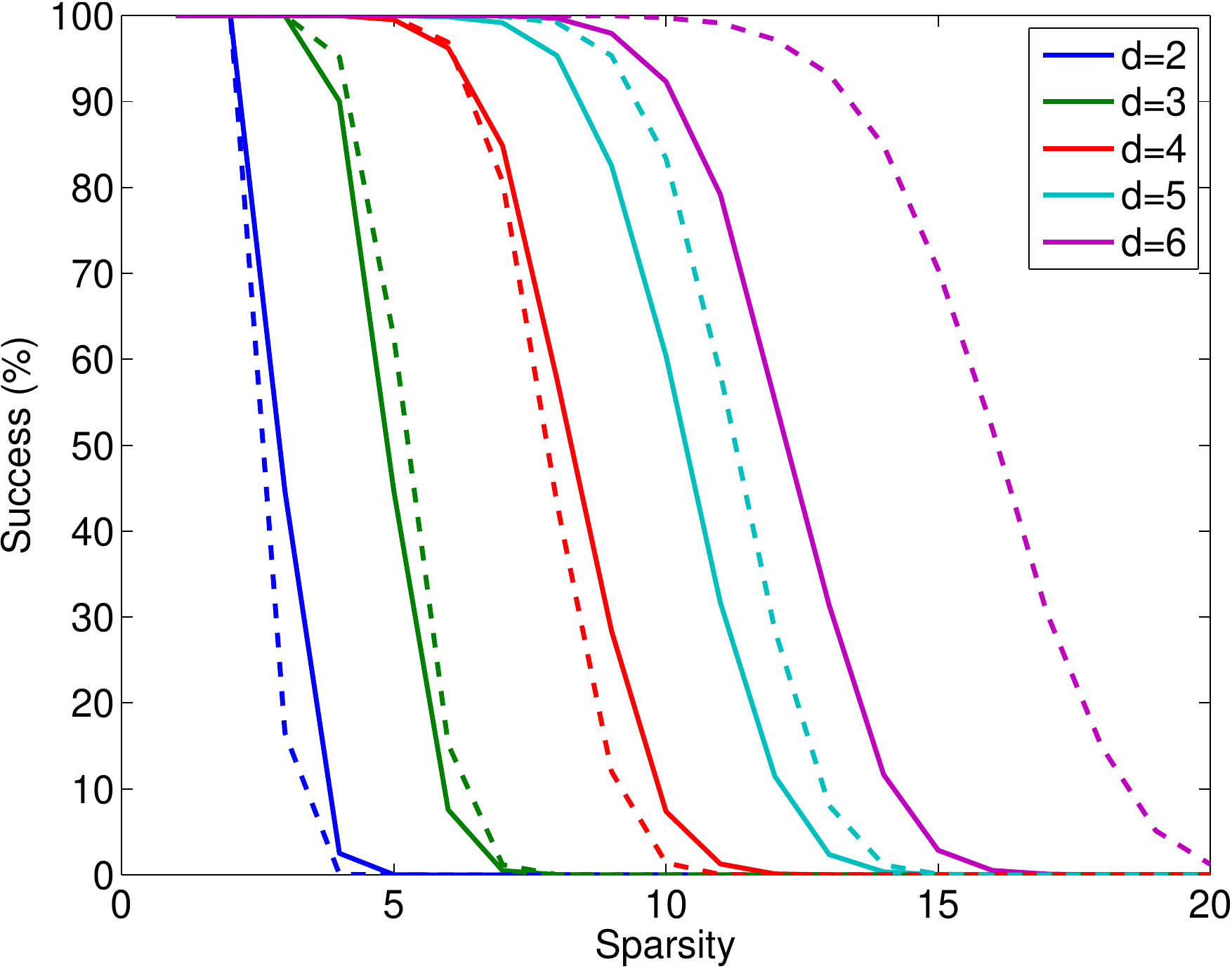}
\caption{Percentage of successfully decoded vectors as a function of
  sparsity level for arrays of size $60\times 60$ (solid) and $120\times
  120$ (dashed). For each sparsity level 10,000 random vectors were
  generated and decoded.}\label{Fig:DecodeSparsity}
\end{figure}

As seen, even when the number of simultaneous pixel firings exceeds
the disjunctness of the group-testing design, it often remains
possible to uniquely decode the resulting codeword. To get more
accurate statistics, we studied the decoding properties of randomly
generated sparse vectors (corresponding to random pixel firing
patterns) with the number of nonzero entries ranging from 1 to 20. For
each sparsity level we decoded 10,000 vectors and summarize the
success rates in Figure~\ref{Fig:DecodeSparsity}. The plots show that
recovery breaks down only gradually once the sparsity exceeds the
disjunctness level of the matrix.

\section{Discussion}

For the adaptation of the group-testing based design in practical
designs, a number of additional issues needs to be addressed. We
discuss some of the major ones below.

\paragraph{Decoding.}

Since the TDCs store time stamps in a memory buffer, we can perform
the decoding process off-line. This is analogous to the operation of a
digital storage oscilloscope. Aside from the obvious benefit of
reducing the amount of circuitry, it has the advantage that we have a
global view of the entire data, not just the current event, or a short
history of previous events. This data can be used to extract more
information from unsuccessfully decoded events by excluding columns
corresponding to pixels that fired shortly before or after the current
event, or by using more sophisticated, but perhaps combinatorial
decoding techniques.

\paragraph{Asynchronicity of the system.}

One challenge in the implementation of the proposed design is to make
sure that the TDCs connected to a single pixel all record the firing
of that pixel in the same TDC interval. In other words, it is
important that codewords are recorded in a single interval and not
spread out over two consecutive intervals. Such a shift can occur due
to differences of travel times of signals between the pixel and the
TDCs as a result of differences in wiring lengths. The appearance of
partial codewords in test vectors complicates decoding and may cause
pixel firings to be masked by others and be missed completely with no
means of detection. 

There are two main approaches of dealing with this problem. The first
approach is to synchronize the logical signal generated by the pixel
using a global clock across the chip to ensure uniform arrival of
signals to the TDC digitizers. Such an approach has been successfully
implemented in clock design trees of CPU chips that maintain clock
skews less than the TDC sampling frequency \cite{RES2001MWCa}. The
second approach is to consider bursts of consecutive events and decode
the union of the corresponding test vectors as a whole. This also
leads to an increase in the expected number of simultaneous firings as
well as a slight decrease in temporal resolution. This approach is
possible only when the photon flux is sufficiently low; when many
consecutive TDC windows contain events, their union simply contains
too many pixel firings to be successfully decoded.

\section*{Acknowledgments}
  E. C., E. vdB., and C. S-L. would like to acknowledge support from the
  NSF via the Waterman Award; from the Broadcom Foundation; from the
  ONR via grant N00014-10-1-0599; and from AFOSR-MURI under grant
  FA9550-09-1-06-43. P.O. would like to acknowledge support through a
  Stanford Interdisciplinary Graduate Student Fellowship.



\begin{thebibliography}{10}

\bibitem{AGO2003AAAa}
S.~Agostinelli et~al.
\newblock Geant4 -- a simulation toolkit.
\newblock {\em Nuclear {I}nstruments and {M}ethods in {P}hysics {R}esearch
  Section {A}}, 506(3):250--303, 2003.

\bibitem{ALO1992Sa}
N.~Alon and J.~H. Spencer.
\newblock {\em The Probabilistic Method}.
\newblock Wiley Series in Discrete Mathematics and Optimization. Wiley, New
  York, 1992.
\newblock (2nd Edition, 2000).

\bibitem{AUL2002LYHa}
B.~F. Aull, A.~H. Loomis, D.~J. Young, R.~M. Heinrichs, B.~J. Felton, P.~J.
  Daniels, and D.~J. Landers.
\newblock Geiger-mode avalanche photodiodes for three-dimensional imaging.
\newblock {\em Lincoln Laboratory Journal}, 13(2):335--350, 2002.

\bibitem{BAL1996BKTa}
D.~J. Balding, W.~J. Bruno, E.~Knill, and D.~C. Torney.
\newblock A comparative survey of non-adaptive pooling designs.
\newblock In T.~P. Speed and M.~S. Waterman, editors, {\em Genetic Mapping and
  {DNA} Sequencing}, pages 133--154. Springer, 1996.

\bibitem{BAR1973a}
Z.~Baranyai.
\newblock On the factorization of exponential sized families of $k$-independent
  sets.
\newblock In {\em Proc. Colloq. Math. Soc. J\'{a}nos B\'{o}lyai, 10th, Infinite
  and Finite Sets}, 1973.

\bibitem{BROxxxxa}
A.~E. Brouwer.
\newblock Bounds for constant weight codes.
\newblock Available online at
  \weburl{http://www.win.tue.nl/~aeb/codes/Andw.html}.

\bibitem{CAS2009SVa}
I.~F. Castro, A.~J. Soares, and J.~F. Veloso.
\newblock Impact of dark counts in low-light level silicon photomultiplier
  multi-readout applications.
\newblock In {\em 2009 IEEE Nuclear Science Symposium Conference Record
  (NSS/MIC)}, pages 1592--1596, November 2009.

\bibitem{CHA1996Ra}
S.~Chaudhuri and J.~Radhakrishnan.
\newblock Deterministic restrictions in circuit complexity.
\newblock In {\em Proceedings of the twenty-eighth annual {ACM} symposium on
  Theory of computing}, STOC '96, pages 30--36, New York, NY, USA, 1996. ACM.

\bibitem{CHE2008Da}
Y.~Cheng and D.-Z. Du.
\newblock New constructions of one- and two-stage pooling designs.
\newblock {\em Journal of Computational Biology}, 15:195--205, 2008.

\bibitem{CLE2003MSa}
A.~E. Clementi, A.~Monti, and R.~Silvestri.
\newblock Distributed broadcast in radio networks of unknown topology.
\newblock {\em Theoretical Computer Science}, 302(1--3):337--364, 2003.

\bibitem{CLI2007EPRa}
R.~Clifford, K.~Efremenko, E.~Porat, and A.~Rothschild.
\newblock $k$-mismatch with don't cares.
\newblock In {\em Proceedings of the 15th annual European conference on
  Algorithms}, ESA'07, pages 151--162, Berlin, Heidelberg, 2007.
  Springer-Verlag.

\bibitem{COL2007Da}
C.~J. Colbourn and J.~H. Dinitz.
\newblock {\em Handbook of Combinatorial Designs}.
\newblock Discrete Mathematics and Its Applications. Chapman \& Hall/CRC, 2007.
\newblock Series Editor: Kenneth H. Rosen.

\bibitem{COR2006Ma}
G.~Cormode and S.~Muthukrishnan.
\newblock Combinatorial algorithms for compressed sensing.
\newblock In P.~Flocchini and L.~Gasieniec, editors, {\em Structural
  Information and Communication Complexity}, volume 4056 of {\em Lecture Notes
  in Computer Science}, pages 280--294. Springer Berlin / Heidelberg, 2006.

\bibitem{DEB2005GVa}
A.~{De Bonis}, L.~G\c{a}sieniec, and U.~Vaccaro.
\newblock Optimal two-stage algorithms for group testing problems.
\newblock {\em {SIAM} Journal on Computing}, 34(5):1253--1270, 2005.

\bibitem{DEB2003Va}
A.~{De Bonis} and U.~Vaccaro.
\newblock Constructions of generalized superimposed codes with applications to
  group testing and conflict resolution in multiple access channels.
\newblock {\em Theoretical Computer Science}, 306(1--3):223--243, 2003.

\bibitem{DOR1943a}
R.~Dorfman.
\newblock The detection of defective members of large populations.
\newblock {\em The Annals of Mathematical Statistics}, 14(4):436--440, 1943.

\bibitem{DU2000Ha}
D.-Z. Du and F.~K. Hwang.
\newblock {\em Combinatorial Group Testing and Its Applications}.
\newblock World Scientific Publishing Company, 2000.

\bibitem{DU2006Ha}
D.~Z. Du and F.~K. Hwang.
\newblock {\em Pooling Designs and Nonadaptive Group Testing: Important Tools
  for {DNA} Sequencing}, volume~18 of {\em Series on Applied Mathematics}.
\newblock World Scientific, New York, 2006.

\bibitem{DYA2000MRa}
A.~G. {D'yachkov}, A.~J. Macula, and V.~V. Rykov.
\newblock New constructions of superimposed codes.
\newblock {\em {IEEE} Transactions on Information Theory}, 46:284--290, 2000.

\bibitem{DYA2000MRb}
A.~G. {D'yachkov}, A.~J. Macular, and V.~V. Rykov.
\newblock {\em New Applications and Results of Superimposed Code Theory Arising
  from the Potentialities of Molecular Biology}, chapter~22, pages 265--282.
\newblock Kluwer Academic Publishers, 2000.

\bibitem{DYA1983Ra}
A.~G. {D'yachkov} and V.~V. Rykov.
\newblock A survey of superimposed code theory.
\newblock {\em Problems of Control and Information Theory}, 14:1--13, 1983.

\bibitem{EPP2007GHa}
D.~Eppstein, M.~T. Goodrich, and D.~S. Hirschberg.
\newblock Improved combinatorial group testing algorithms for real-world
  problem sizes.
\newblock {\em {SIAM} Journal on Computing}, 36(5):1360--1375, 2007.

\bibitem{ERD1982FFa}
P.~Erd\"{o}s, P.~Frankl, and Z.~F\"{u}redi.
\newblock Families of finite sets in which no set is covered by the union of
  two others.
\newblock {\em Journal of Combinatorial Theory, Series {A}}, 33(2):158--166,
  1982.

\bibitem{ERD1985FFa}
P.~Erd\"{o}s, P.~Frankl, and Z.~F\"{u}redi.
\newblock Families of finite sets in which no set is covered by the union of
  $r$ others.
\newblock {\em Israel J. Math.}, 51:79--89, 1985.

\bibitem{ERD1974La}
P.~Erd\"{o}s and L.~Lov\'{a}sz.
\newblock Infinite and finite sets.
\newblock {\em Colloq. Math. Soc. {J\~{a}nos} {Bolyai}}, 10:609--627, 1974.

\bibitem{FAR1997KKMa}
M.~Farach, S.~Kannan, E.~Knill, and S.~Muthukrishnan.
\newblock Group testing problems with sequences in experimental molecular
  biology.
\newblock In {\em Proceedings of Compression and Complexity of Sequences 1997},
  pages 357--367, June 1997.

\bibitem{FRA2009PDGa}
T.~Frach, G.~Prescher, C.~Degenhardt, R.~de~Gruyter, A.~Schmitz, and
  R.~Ballizany.
\newblock The digital silicon photomultiplier -- principle of operation and
  intrinsic detector performance.
\newblock In {\em 2009 IEEE Nuclear Science Symposium Conference Record
  (NSS/MIC)}, pages 1959--1965, October 2009.

\bibitem{FUR1996a}
Z.~F\"{u}redi.
\newblock On $r$-cover-free families.
\newblock {\em Journal of Combinatorial Theory, Series {A}}, 73:172--173, 1996.

\bibitem{GORxxxxa}
D.~Gordon et~al.
\newblock {La Jolla Covering Repository}.
\newblock Available online at \weburl{http://www.ccrwest.org/cover.html}.

\bibitem{GRAxxxxa}
M.~Grassl.
\newblock Bounds on the minimum distance of linear codes and quantum codes.
\newblock Available online at \weburl{http://www.codetables.de}, 2007.

\bibitem{HEN2009GZSa}
D.~Henseler, R.~Grazioso, N.~Zhang, and M.~Schmand.
\newblock {SiPM} performance in {PET} applications: An experimental and
  theoretical analysis.
\newblock In {\em Nuclear Science Symposium Conference Record (NSS/MIC), 2009
  IEEE}, pages 1941--1948, November 2009.

\bibitem{HWA1987Sa}
F.~K. Hwang and V.~T. S\'{o}s.
\newblock Non adaptive hypergeometric group testing.
\newblock {\em Studia Sci. Math. Hungarica}, 22:257--263, 1987.

\bibitem{IND1997a}
P.~Indyk.
\newblock Deterministic superimposed coding with applications to pattern
  matching.
\newblock In {\em Proceedings of the 38th Annual Symposium on Foundations of
  Computer Science}, pages 127--136, Washington, DC, USA, 1997. IEEE Computer
  Society.

\bibitem{ISA2010PBHa}
S.~Isaak, M.~C. Pitter, S.~Bull, and I.~Harrison.
\newblock Design and characterisation of $16\times 1$ parallel outputs {SPAD}
  array in 0.18 um {CMOS} technology.
\newblock In {\em 2010 IEEE Asia Pacific Conference on Circuits and Systems
  (APCCAS)}, pages 979--982, December 2010.

\bibitem{JOH1963a}
S.~M. Johnson.
\newblock Improved asymptotic bounds for error-correcting codes.
\newblock {\em {IEEE} Transactions on Information Theory}, 9(4):198--205, 1963.

\bibitem{KAU1964Sa}
W.~H. Kautz and R.~C. Singleton.
\newblock Nonrandom binary superimposed codes.
\newblock {\em {IEEE} Transactions on Information Theory}, 10(4):363--377,
  October 1964.

\bibitem{KNI1998BTa}
E.~Knill, W.~J. Bruno, and D.~C. Torney.
\newblock Non-adaptive group testing in the presence of errors.
\newblock {\em Discrete Applied Mathematics}, 88(1--3):261--290, November 1998.

\bibitem{MAC1997a}
A.~J. Macula.
\newblock Error-correcting nonadaptive group testing with $d^e$-disjunct
  matrices.
\newblock {\em Discrete Applied Mathematics}, 80(2--3):217--222, December 1997.

\bibitem{MAC1981Sa}
F.~J. Mac{W}illiams and N.~J.~A. Sloane.
\newblock {\em The Theory of Error-Correcting Codes}.
\newblock North-Holland Publishing Company, Amsterdam, 1977.
\newblock (3rd printing, 1983).

\bibitem{MAR2010TZTa}
B.~Markovic, A.~Tosi, F.~Zappa, and S.~Tisa.
\newblock Smart-pixel with {SPAD} detector and time-to-digital converter for
  time-correlated single photon counting.
\newblock In {\em 2010 23rd Annual Meeting of the IEEE Photonics Society},
  pages 181--182, November 2010.

\bibitem{NGO2000Da}
H.~Q. Ngo and D.-Z. Du.
\newblock A survey on combinatorial group testing algorithms with applications
  to {DNA} library screening.
\newblock In {\em Discrete Mathematical Problems with Medical Applications},
  volume~55 of {\em {DIMACS} Ser. Discrete Math. Theoret. Comput. Sci.}, pages
  171--182, Providence, RI, 2000. American Mathematical Society.

\bibitem{POR2008Ra}
E.~Porat and A.~Rothschild.
\newblock Explicit non-adaptive combinatorial group testing schemes.
\newblock In L.~Aceto, I.~Damg{\aa}rd, L.~Goldberg, M.~Halld{\'o}rsson,
  A.~Ing{\'o}lfsd{\'o}ttir, and I.~Walukiewicz, editors, {\em Automata,
  Languages and Programming}, volume 5125 of {\em Lecture Notes in Computer
  Science}, pages 748--759. Springer Berlin / Heidelberg, 2008.

\bibitem{REN2006a}
D.~Renker.
\newblock Geiger-mode avalanche photodiodes, history, properties and problems.
\newblock {\em Nuclear Instruments and Methods in Physics Research Section A:
  Accelerators, Spectrometers, Detectors and Associated Equipment},
  567(1):48--56, 2006.

\bibitem{RES2001MWCa}
P.~J. Restle, T.~G. McNamara, D.~A. Webber, P.~J. Camporese, K.~F. Eng, K.~A.
  Jenkins, S.~Member, D.~H. Allen, M.~J. Rohn, M.~P. Quaranta, D.~W. Boerstler,
  C.~J. Alpert, C.~A. Carter, R.~N. Bailey, J.~G. Petrovick, B.~L. Krauter, and
  B.~D. McCredie.
\newblock A clock distribution network for microprocessors.
\newblock {\em IEEE Journal of Solid-State Circuits}, 36(5):792--799, May 2001.

\bibitem{RIC1998Ca}
L.~Riccio and C.~J. Colbourn.
\newblock An upper bound for disjunct matrices.
\newblock Preprint, University of Vermont, 1998.

\bibitem{ROD2004MOPa}
P.~Rodrigues, R.~Moura, C.~{Ortig\~{a}o}, L.~Peralta, M.~G. Pia, A.~Trindade,
  and J.~Varela.
\newblock Geant4 applications and developments for medical physics experiments.
\newblock {\em IEEE Transactions on Nuclear Science}, 51(4):1412--1419, August
  2004.

\bibitem{RUS1994a}
M.~Ruszink\'{o}.
\newblock On the upper bound of the size of the $r$-cover-free families.
\newblock {\em Journal of Combinatorial Theory, Series {A}}, 66:302--310, 1994.

\bibitem{SEF2007a}
F.~Sefkow.
\newblock The {CALICE} tile hadron calorimeter prototype with {SiPM} read-out:
  Design, construction and first test beam results.
\newblock In {\em Nuclear Science Symposium Conference Record, 2007. NSS '07.
  IEEE}, volume~1, pages 259--263, November 2007.

\bibitem{SIN1964a}
R.~C. Singleton.
\newblock Maximum distance q-nary codes.
\newblock {\em {IEEE} Transactions on Information Theory}, 10(2):116--118,
  April 1964.

\bibitem{SOB1959Ga}
M.~Sobel and P.~A. Groll.
\newblock Group testing to eliminate efficiently all defectives in a binomial
  sample.
\newblock {\em Bell Syst. Tech. J.}, 38:1179--1252, 1959.

\bibitem{Software:SAGE}
W.~A. Stein et~al.
\newblock {\em {S}age {M}athematics {S}oftware ({V}ersion 4.7.2)}.
\newblock The {Sage} Development Team, 2012.
\newblock \weburl{http://www.sagemath.org}.

\bibitem{TES2007DMNa}
M.~Teshima, B.~Dolgoshein, R.~Mirzoyan, J.~Nincovic, and E.~Popova.
\newblock {SiPM} development for astroparticle physics applications.
\newblock In {\em Proceedings of the 30th International Cosmic Ray Conference},
  volume~5, pages 985--988, 2007.

\bibitem{WOL1985a}
J.~K. Wolf.
\newblock Born again group testing: Multiaccess communications.
\newblock {\em {IEEE} Transactions on Information Theory}, 31(2):185--191,
  March 1985.

\bibitem{YEH2002a}
H.-G. Yeh.
\newblock $d$-{D}isjunct matrices: Bounds and {L}ov\'{a}sz local lemma.
\newblock {\em Discrete Mathematics}, 253(1--3):97--107, June 2002.

\end{thebibliography}

\end{document}